\documentclass[12pt]{article}
\usepackage{graphicx}
\newcommand{\mvector}{\ensuremath{0.882}}
\newcommand{\mzero}{\ensuremath{1.174}}

\newcommand{\sla}[1]{{#1}\!\!\!/}

\newcommand{\spac}{\ensuremath{\ }}
\newcommand{\kl}{\ensuremath{K_{\ell 3}\spac}}

\newcommand{\kln}{\ensuremath{K^0_{\ell 3}\spac}}
\newcommand{\ken}{\ensuremath{K^0_{e 3}\spac}}
\newcommand{\kmn}{\ensuremath{K^0_{\mu 3}\spac}}
\newcommand{\klc}{\ensuremath{K^+_{\ell 3}\spac}}
\newcommand{\kec}{\ensuremath{K^+_{e 3}\spac}}

\newcommand{\dtot}{\ensuremath{\delta^{\ell}_{T}}}
\newcommand{\dtote}{\ensuremath{\delta^{e}_{T}}}
\newcommand{\dtotm}{\ensuremath{\delta^{\mu}_{T}}}
\newcommand{\dshort}{\ensuremath{\delta_{SD}}}
\newcommand{\dlong}{\ensuremath{\delta_{LD}}}
\newcommand{\dlongl}{\ensuremath{\delta^{\ell}_{LD}}}
\newcommand{\dlonge}{\ensuremath{\delta^{e}_{LD}}}
\newcommand{\dlongm}{\ensuremath{\delta^{\mu}_{LD}}}
\newcommand{\hdlongl}{\ensuremath{\hat{\delta}^{\ell}_{LD}}}
\newcommand{\hdlonge}{\ensuremath{\hat{\delta}^{e}_{LD}}}
\newcommand{\hdlongm}{\ensuremath{\hat{\delta}^{\mu}_{LD}}}
\newcommand{\mlsq}{\ensuremath{m^2_{\ell}}}

\newcommand{\mpisq}{\ensuremath{m^2_{\pi}}}
\newcommand{\mksq}{\ensuremath{m^2_{K}}}
\newcommand{\mplussq}{\ensuremath{m^2_{+}}}
\newcommand{\mzerosq}{\ensuremath{m^2_{0}}}
\newcommand{\kpln}{\ensuremath{K^0\pi^-\ell^+\nu_{\ell}}}
\newcommand{\kplng}{\ensuremath{K^0\pi^-\ell^+\nu_{\ell}\gamma}}
\newcommand{\kplngg}{\ensuremath{K^0\pi^-\ell^+\nu_{\ell}\gamma\gamma}}
\newcommand{\klng}{\ensuremath{K^0_{\ell 3\gamma}}}
\newcommand{\xpt}{ChPT}

\newcommand{\dip}{\ensuremath{D_i^{(+)}}}
\newcommand{\dzp}{\ensuremath{D_0^{(+)}}}
\newcommand{\dop}{\ensuremath{D_1^{(+)}}}
\newcommand{\dtwp}{\ensuremath{D_2^{(+)}}}
\newcommand{\dthp}{\ensuremath{D_3^{(+)}}}
\newcommand{\dio}{\ensuremath{D_i^{(0)}}}

\newcommand{\mkmmpi}{\ensuremath{(\mksq-\mpisq )}}

\begin{document}

\centerline{\bf Radiative Corrections in \kln Decays}
\medskip

\centerline{Troy C. Andre\footnote{\tt troy@harimad.com}}
\medskip

\centerline{\it Enrico Fermi Institute and Department of Physics}
\centerline{\it University of Chicago, Chicago, IL 60637}

\date{\today}

\leftline{\bf Abstract}

\begin{quote}
We calculate the long-distance radiative corrections \dlonge\spac and
\dlongm\spac to the $K^0\rightarrow \pi^{-} e^{+} \nu_{e}$ and the
$K^0\rightarrow \pi^{-} \mu^{+} \nu_{\mu}$ decay rates.  This analysis
includes contributions to the long-distance radiative corrections from
outside the kinematically-allowed three-body Dalitz region and tests
the sensitivity of the radiative corrections to the hadronic
$K$--$\pi$ form factors.  A program, {\tt KLOR}, was written to
numerically evaluate the radiative corrections and to generate Monte
Carlo events for experimental acceptance studies.  The \ken and the
\kmn long-distance radiative correction parameters are determined to
be $(1.3\pm 0.3)\%$ and $(1.9\pm 0.3)\%$, respectively.  We also
present predictions for the fraction of radiative \kln events
satisfying various requirements on final-state photon kinematics.
\end{quote}


\section{\label{sec:intro}Introduction}
In the Standard Model, the strength and the pattern of weak
interactions between up- and down-type quarks are encoded in a
$3\times 3$ unitarity matrix: the Cabibbo-Kobayashi-Maskawa (CKM)
matrix~\cite{CKM}.  Years of experimental and theoretical research
have gone into the determination of the elements of this matrix.
These measurements, when combined, test the unitarity of the CKM
matrix and thereby determine if the Standard Model can fully explain
quark mixings.  Moreover, measurements of the CKM matrix elements can
be used to limit proposed extensions of the Standard Model.

The $V_{us}$ element of the CKM matrix is of particular interest
because CKM unitarity is best tested in the first row of the matrix.
$|V_{us}|$ is measured using three-body semileptonic decays of the $K$
meson, hyperon decays, and tau decays.  Analysis of the semileptonic
kaon decays, denoted $K_{\ell 3}$, leads to the most accurate measure
of $|V_{us}|$.  Both the charged ($K^+_{\ell 3}$) and the neutral
($K^0_{\ell 3}$) kaon decays provide independent measurements of $|V_{us}|$.

Up to 2004, analysis of the \kec and the \ken decay modes yielded $|V_{us}| =
0.2200\pm 0.0026$~\cite{Eidelman:2004wy}.  The measurements in the \kln and
\klc modes were in agreement at the 1\% level; however, they were inconsistent
with the unitarity of the CKM matrix at the $2\sigma$ level.  This situation
soon changed dramatically.  A measurement \cite{Sher:2003fb} of $|V_{us}|$ from
the \kec mode obtained a value of $|V_{us}|$ that is $\sim 3\%$ higher and
consistent with CKM unitarity.  Results from KTeV~\cite{Alexopoulos:2004sw,%
Alexopoulos:2004sx,Alexopoulos:2004sy} indicated a value of $|V_{us}|$,
measured in the \ken and \kmn modes, that is also consistent with CKM
unitarity.  The precision of the measurement required an understanding
of physical effects at better than the percent level.  Although radiative
corrections to these processes had been calculated within chiral perturbation
theory \cite{Bijnens,Cirigliano,Bytev:2002nx}, the KTeV experiment measured the
angular distributions of radiated photons, which were not modeled satisfactorily
by currently available treatments \cite{Barberio:1993qi}. The present paper was
written in response to that need.  Further measurements of $|V_{us}|$ were
performed by the IHEP \cite{Yushchenko:2004zs}, CERN NA48 \cite{Lai:2004bt},
and KLOE \cite{Ambrosino:2006gn} Collaborations, leading to a current world
average \cite{Yao:2006px} of $|V_{us}| = 0.2257 \pm 0.0021$.  The KLOE
Collaboration utilized radiative corrections documented in Ref.\
\cite{Gatti:2005kw}.  A concise account of the present status of radiative
corrections to \kln decays is given in Ref.\ \cite{Cirigliano:2006ix}.

In this paper, we study radiative corrections to the \ken and the \kmn
decay modes.  The total radiative correction to the \kln decay rate is
described by a parameter \dtot.  This parameter is the sum of
two contributions: the long-distance (\dlongl) and the short-distance
(\dshort) components.  The short-distance parameter is
known~\cite{marciano} and we calculate the long-distance corrections
using a phenomenological model originally used in
Refs.~\cite{Ginsberg:1966, Ginsberg:jh, Ginsberg:pz, Ginsberg:vy,
Becherrawy:1970, Fearing:zz, Fischbach:qw}.  Using the
phenomenological model, we test the assumption that the variation of
the hadronic $K$-$\pi$ form factors is negligible.  Single-photon
radiation from all \kln decays are included in our calculation of the
long-distance radiative correction.  A program, named {\tt KLOR}, was
written to numerically evaluate the radiative corrections and to
generate Monte Carlo events.  Monte Carlo events may be used by
experimentalists to understand detection efficiency.  For example, the
detection efficiency of \kln decays may be affected by the conversion
of a radiated photon to an electron--positron pair.

We also determine the fraction of radiative \kln events (normalized by
the \kln decay rate) that satisfy various constraints on the radiated
photon's energy and its proximity to the charged lepton.  These
``radiative fractions'' are interesting since they can be checked by
experiment.  Related studies~\cite{Bijnens,Cirigliano,Bytev:2002nx}
calculate the radiative corrections using chiral perturbation theory (\xpt).

The paper is organized as follows.  In Section~\ref{sec:model_nonrad},
we review relevant properties of the phenomenological model for the
lowest-order contribution to the \kln decay.  In
Section~\ref{sec:model_rad}, we extend that model to include radiative
corrections to the \kln decay.  In Section~\ref{sec:numerical_method},
we discuss the numerical method used to calculate the radiative
correction.  In Section~\ref{sec:results}, we present results from our
calculation of the radiative corrections, while
Section~\ref{sec:conclusion} concludes.  In Appendix~\ref{appendix:a},
we define the relevant loop integrals for this analysis and in
Appendix~\ref{appendix:b}, equations used to calculate the virtual
matrix element are presented.  In Appendix~\ref{appendix:c}, we
compare our Monte Carlo generator, {\tt KLOR}, to another Monte Carlo
generator that can be used to approximate radiative corrections in
\kln decays.

\section{Phenomenological Model of \kln Decays}\label{sec:model_nonrad}
The \kln decay mode and associated kinematics are described by the
equation
\begin{equation}\label{eqn:kl3_kin}
K^0(p_K) \rightarrow \pi^-(p_{\pi}) \; \ell^+(p_{\ell}) \; \nu_{\ell}(p_{\nu}),
\end{equation}
where $\ell^+$ ($=$ $e^+$ or $\mu^+$) denotes the charged lepton,
$\nu_{\ell}$ is the $\ell$ neutrino, and $p_X$ is the four-momentum of
particle `$X$' (= $K^0$, $\pi^-$, $\ell^+$, or $\nu_{\ell}$).
Conventionally, the lowest-order contribution to the \kln decay is
studied using a phenomenological model~\cite{Chounet}.  Using this
model and the momentum parametrization of Eq.~(\ref{eqn:kl3_kin}), the
lowest-order (Born) transition matrix element for the \kln decay
is~\cite{footnote:general,Chizhov,tesarek}
\begin{equation}\label{eqn:born_me}
{\cal M}^B = \sqrt{2} G_F V_{us} \bar{u}(p_{\nu})\gamma^{\mu} P_L v(p_{\ell})
[ f_+(t) (p_K+p_{\pi})_{\mu} + (p_K-p_{\pi})_{\mu} f_-(t) ],
\end{equation}
where $G_F$ ($=1.16639\times 10^{-5}$ GeV) is the Fermi coupling
constant, $V_{us}$ is the $u$--$s$ coupling in the CKM matrix, $P_L =
\frac{1}{2}(1-\gamma^5)$ is the left-handed projection operator, and
$t=(p_K-p_{\pi})^2$.  The Mandelstam $t$ variable is the square of the
momentum transfer to the $W^{+}$ boson ($W^{+}\rightarrow \ell^+\nu$).
In Eq.~(\ref{eqn:born_me}), the functions $f_{\pm}(t)$ are the
$K$--$\pi$ hadronic form factors; the form factors arise from the
contraction of the quark current with the on-shell kaon and pion,
\begin{equation}
f_+(t)(p_K+p_{\pi})^{\mu} + f_-(t)(p_K-p_{\pi})^{\mu}
= \langle \pi^-(p_{\pi})|\bar{s}\gamma^{\mu}P_L u|K^0(p_{K})\rangle.
\end{equation}

\subsection{$K$--$\pi$ Hadronic Form Factors}\label{sec:form_factor}
Before proceeding further, the structure and alternate definitions of
the form factors are discussed.  As shown in Eq.~(\ref{eqn:born_me}),
the hadronic form factors only depend on $t$.  This simple functional
dependence is guaranteed by local creation of the lepton pair.  Time
reversal invariance of the \kln decay ensures that the form factors
are relatively real.

Instead of using the set of form factors \{$f_+$, $f_-$\}, an
alternate form factor set \{$f_+$, $f_0$\} is typically used.  The
$f_+(t)$ and $f_0(t)$ form factors correspond to the vector $1^-$ and
the scalar $0^+$ exchange amplitudes, respectively.  This description
of the $K$--$\pi$ interaction is inspired by a vector dominance model
(VDM)~\cite{VDM}.  The \{$f_+$, $f_-$\} form factor set is related to
the \{$f_+$, $f_0$\} form factor set by the relation
\begin{equation}\label{eqn:f0}
f_0(t) = f_+(t) + \frac{t}{m^2_K - m^2_{\pi}}f_-(t),
\end{equation}
where $m_K$ and $m_{\pi}$ are the $K^0$ and $\pi^{-}$ masses, respectively.

Normalized versions of the form factors $f_+(t)$ and $f_0(t)$ are
often analyzed.  The {\it normalized form factors} are defined by the
equation, $\hat{f}_{+,\, 0}(t) = f_{+,\, 0}(t)/f_{+,\, 0}(0)$.
Requiring $f_-(t)$ to be finite as $t \rightarrow 0$ implies that
$f_+(0) = f_0(0)$; therefore, $f_+(0)$ is the normalization factor
common to both $f_+(t)$ and $f_0(t)$.  We will explicitly write our
matrix elements in terms of $\hat{f}_{+,\, 0}(t)$ and $f_+(0)$.

In this paper, we consider three form factor models: the ``linear
model,'' the ``quadratic model,'' and the ``pole model.''  The
linear and the quadratic form factor models capture the first- and
second-order dependence of the form factor on the momentum transfer,
$t$.  The normalized linear and quadratic form factor models are
described by the equations
\begin{eqnarray}\label{eqn:linear_ff}
\hat{f}^{\,{\rm lin}}_{+,\, 0}(t)
 & = & 1 + \lambda_{+,\,0} \frac{t}{m^2_{\pi}}, \\
\hat{f}^{\,{\rm quad}}_{+,\, 0}(t)
 & = & 1 + \lambda^{\prime}_{+,\,0} \frac{t}{m^2_{\pi}} 
 + \frac{1}{2} \lambda^{\prime\prime}_{+,\,0} \frac{t^2}{m^4_{\pi}}, \nonumber
\end{eqnarray}
where the $\lambda$s are model parameters that are measurable by
experiment.  Since the linear and quadratic form factor models are
Taylor-series approximations of the ``true'' form factors, they do not
describe the higher-order or the large $t$- dependence of the
$K$--$\pi$ form factors.

Unlike the linear and quadratic form factor models, the pole model
describes form factor $t$-dependence beyond second-order and for large
momentum transfers.  The pole model supposes that the behavior of the
form factors $f_+(t)$ and $f_0(t)$ are dominated by the exchange of
the lightest vector and scalar $K^*$ mesons.  The pole model expects
the $K^*$ mass poles to be $m_+ \approx 890$ MeV/c$^2$ and $m_0 >
m_+$.  In this paper, we consider a simple pole model in which the
form factors are given by the equation
\begin{equation}\label{eqn:pole_ff}
\hat{f}^{{\rm pole}}_{+,\, 0}(t) = \frac{m^2_{+,\,0}}{m^2_{+,\,0} - t}.
\end{equation}
Modified pole-model schemes (which are not considered in this paper)
may include the effects of the $K^*$ width or the effects of mixing
with other vector-meson poles.

\subsection{Born Decay Rate}\label{sec:gamma_born}
Using the matrix element for the \kln decay mode [see
Eq.~(\ref{eqn:born_me})], the Born contribution to the semileptonic
decay rate is
\begin{equation}\label{eqn:bornrate}
\Gamma^{(0)}_{K\ell 3} = \frac{G_F^2 m_K^5}{192\pi^3} \, 
 {|V_{us}|}^2 f^2_+(0) \, {\mathcal I}^{\ell}_{(0)},
\end{equation}
where $\ell$ ($=e$ or $\mu$), $f_+(0)$ is the form factor
normalization constant and ${\mathcal I}^{\ell}_{(0)}$ is the
lowest-order phase space integral.  The phase space integral is given
by the expression~\cite{Leutwyler:1984je}
\begin{equation}\label{eqn:psintegral0}
{\mathcal I}^{\ell}_{(0)} = \frac{1}{m_K^8}\int_{m_{\ell}^2}^{t_{\rm max}} dt~
 {\left[{\lambda(t,m_K^2,m_{\pi}^2)}\right]}^{\frac{3}{2}}~F(t)
  \left(1+\frac{m_{\ell}^2}{2t} \right) \left(1-\frac{m_{\ell}^2}{t}\right)^2,
\end{equation}
where $m_{\ell}$ ($\ell = e$ or $\mu$) is the mass of the charged
lepton, $t_{\rm max} = (m_K^2-m_{\pi}^2)^2$,
\begin{equation}\label{eqn:psintegral_argument}
F(t) = \hat{f}^2_+(t)+\frac{3m^2_{\ell}(m_K^2-m_{\pi}^2)}
  {\lambda(t,m_K^2,m_{\pi}^2) \, (2t+m_{\ell}^2)}\hat{f}^2_0(t),
\end{equation}
and $\lambda(a,b,c) = a^2 + b^2 + c^2 - 2ab - 2bc - 2ca$.  From
Eq.~(\ref{eqn:psintegral_argument}), one observes that the
semileptonic decay rate depends on the $K$--$\pi$ form factors.

\section{Radiative Corrections to \kln Decays}\label{sec:model_rad}
The \kln decay rate receives corrections from photon emission and
exchange.  Including first-order radiative corrections, the $K^0_{\ell
3}$ decay rate is
\begin{equation}\label{eqn:tot_rate}
\Gamma_{K\ell 3} = \Gamma^{(0)}_{K\ell 3} + \Gamma^{(1)}_{K\ell 3}
 = \frac{G_F^2 m_K^5}{192\pi^3} {|V_{us}|}^2 f^2_+(0)
 \left( 1 + \dtot \right) {\mathcal I}^{\ell}_{(0)},
\end{equation}
where \dtot\spac is the total radiative correction to the Born \kln
decay rate, $f_+(0)$ is the form factor normalization constant, and
${\mathcal I}^{\ell}_{(0)}$ is the lowest-order phase-space integral
given in Eq.~(\ref{eqn:psintegral0}).  The radiative correction,
\dtot, is the sum of two components~\cite{footnote:delta}: the
short-distance (\dshort) and long-distance (\dlongl) correction.
Including the effects of quantum chromodynamics (QCD), the short
distance radiative correction is~\cite{marciano}
\begin{eqnarray}\label{eqn:short_distance}
\dshort & = & \frac{\alpha}{\pi} \left(1- \frac{\alpha_s}{\pi}\right) 
              \ln\frac{m_Z^2}{\Lambda^2},
\end{eqnarray}
where $\alpha$ is the fine structure constant, $\alpha_s =
\alpha_s(M^2_Z)$ is the strong coupling constant at the $Z^0$ mass
scale, $m_Z$ is the mass of the $Z^0$ boson, and $\Lambda$ is the
low-energy cutoff.  The low-energy cutoff is the hadronic mass scale
used as a boundary between the short- and long-distance loop
corrections.  The specific cutoff scale is somewhat arbitrary and it
usually depends on the limitations of the long-distance model; two
common cutoff scales selected are the proton mass, $m_p$, and the
$\rho$ meson mass, $m_{\rho}$.  The effect of the low-energy cutoff
scale is addressed in Sec.~\ref{sec:results}.

The long-distance radiative correction is obtained from a model of the
interaction between the photon, the hadrons, and the charged lepton.
Original studies of the long-distance radiative correction used an
extension of the phenomenological model described in
Sec.~\ref{sec:model_nonrad}.  Application of this model to the \kln
and the \klc decays are in Refs.~\cite{Ginsberg:pz,Ginsberg:vy} and
Refs.~\cite{Ginsberg:1966,Ginsberg:jh,Becherrawy:1970}, respectively.
More recent studies~\cite{Bijnens,Cirigliano,Bytev:2002nx} calculate
the long-distance contribution to the radiative correction using \xpt.

\subsection{Long Distance Radiative Correction}
We revisit the phenomenological model for radiative corrections to
the \kln decay mode.  In this section, we outline the assumptions of
the previous studies, discuss our modification of these assumptions,
and calculate the matrix elements that contribute to the \kln
radiative correction.

In Refs.~\cite{Ginsberg:pz,Ginsberg:vy}, the long-distance radiative
corrections to the \kln decay modes are computed under the following
assumptions:
\begin{itemize}
\item[${\mathcal A}1.$]{Radiative effects are accurately described by
first-order perturbation theory in $\alpha$.}
\item[${\mathcal A}2.$]{The phenomenological hadronic model [see
Sec.~\ref{sec:model_nonrad}] characterizes the weak $K$--$\pi$
vertex.}
\item[${\mathcal A}3.$]{The $K$--$\pi$ form factors are not modified
by the presence of radiative effects.}
\item[${\mathcal A}4.$]{The effect of the pion form factor is
negligible.}
\item[${\mathcal A}5.$]{The $t$ dependence of the $K$--$\pi$ form
factors produces a negligible effect on the radiative corrections.}
\item[${\mathcal A}6.$]{The experimental apparatus used to detect the
radiative \kln decays (\klng) only detects the charged lepton and the
pion; the neutrino and the {\it photon} are undetected.}
\end{itemize}
For our calculation, we retain assumptions ${\mathcal A}1$--${\mathcal
A}4$, we test assumption ${\mathcal A}5$, and we remove assumption
${\mathcal A}6$ since radiated photons are detected and influence
detection efficiencies.

\begin{figure}[h]
\begin{center}
\includegraphics[scale=0.6]{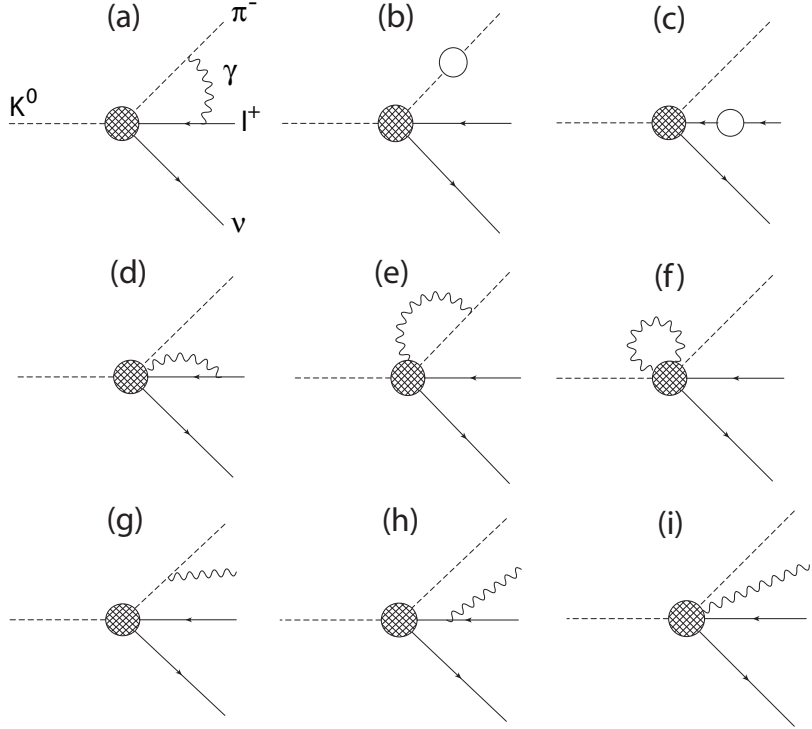}
\end{center}
\caption{\label{fig:feynman} Feynman diagrams for the first-order
radiative corrections to the $K^0_{\ell 3}$ decay mode.  Diagrams
(a)--(f) are the {\it virtual} corrections while diagrams (g)--(i) are
the {\it inner-bremsstrahlung} corrections.  The open circles in
diagram (b) and (c) denote the self-energy correction to the pion and
lepton wavefunctions, respectively.}
\end{figure}

Under the first assumption, ${\mathcal A}1$, the Feynman diagrams
contributing to the first-order radiative corrections are shown in
Fig.~\ref{fig:feynman}.  The first-order radiative corrections consist
of two classes of diagrams: virtual and inner-bremsstrahlung.  Virtual
diagrams, Fig.~\ref{fig:feynman}(a)--(f), involve the emission of a
photon by the effective vertex, the charged lepton, or the pion and
subsequent absorption by the effective vertex, the charged lepton, or
the pion.  The inner-bremsstrahlung diagrams,
Fig.~\ref{fig:feynman}(g)--(i), involve the emission of a real photon
by the pion, the charged lepton, or the effective vertex.

Though higher-order corrections in $\alpha$, electromagnetic
corrections to strong-interaction graphs, and weak interactions also
contribute to the radiative correction, their effects are assumed to
be small.  For studies that model the effects of electromagnetic
corrections to strong-interaction graphs, we refer the reader to the
\xpt\spac studies~\cite{Bijnens,Cirigliano,Bytev:2002nx}.

To calculate the first-order radiative corrections, we need an
expression for the ``\kpln'' effective vertex when at least one of the
particle legs is {\it off-shell} [see Fig.~\ref{fig:feynman}(a),
(g)--(h)].  Under ${\mathcal A}2$, the {\it off-shell} effective
vertex is obtained from the {\it on-shell} phenomenological model.
Therefore, the \kpln\spac vertex is
\begin{equation}\label{eqn:effective_vertex}
V^{K\pi\ell\nu} = i\sqrt{2}G_F V_{us} f_+(0) P_R
 \left[ (\sla{p}^{*}_K + \sla{p}^{*}_{\pi})\hat{f}_+(t^*) 
 +(\sla{p}^{*}_K - \sla{p}^{*}_{\pi})\hat{f}_-(t^*) \right],
\end{equation}
where $p^*_K$ and $p^*_{\pi}$ are the (potentially) off-shell $K^0$
and $\pi^-$ four-momenta, $t^*=(p^*_K - p^*_{\pi})^2$, and the vector
$\sla{p}$ denotes the contraction of the four-vector $p^{\mu}$ with
the Dirac matrices, $\gamma^{\mu}$.

In general, one expects Eq.~(\ref{eqn:effective_vertex}) to be an
approximation of the {\it true} off-shell vertex.  One potential
correction to this vertex is a modification of the $K$--$\pi$ form
factors.  The form factors for the off-shell vertex may depend on
kinematical quantities beyond $t$.  In fact, using \xpt,
Ref.~\cite{Cirigliano} finds that the form factors depend on the
additional variable $s=(p_{\pi}+p_{\ell})^2$ (e.g. $f_{\pm}(t,s)$).
Since the emission of low-energy photons is expected to dominate the
radiative corrections, we assume (${\mathcal A}3$) that the variation
of the form factors $f_{+,\,0}$ is captured by simple $t$-dependence.

The expression for the off-shell \kpln\spac vertex may be simplified
by using energy-momentum conservation and the Dirac equation for the
neutrino.  The simplified expression for this vertex is
\begin{equation}\label{eqn:effective_vertex_compact}
V^{K\pi\ell\nu} = i\sqrt{2}G_F V_{us} f_+(0) P_R
 \left[ 2\sla{p}^*_K \hat{f}_+(t^*) 
 -\sla{p}^*_{\ell} \hat{f}_2(t^*) \right],
\end{equation}
where $\hat{f}_2(t^*) = \hat{f}_+(t^*) - \hat{f}_-(t^*)$ is the
difference between the $\hat{f}_+$ and $\hat{f}_-$ form factors.

We assume that the electromagnetic interaction with the pion is
point-like (${\mathcal A}4$).  Deviations from the point-like
assumption are captured by the introduction of pion form factors.  For
experimental studies of the strength of pion form factors consult
Ref.~\cite{Yao:2006px}; in this study, we assume pion form factors are
negligible for current experimental sensitivities.  Previous
studies~\cite{Becherrawy:1970} of radiative corrections to the \kec
decay mode include the effect of the pion form factors.  Though the
long-distance radiative correction to the \kec mode from
Ref.~\cite{Becherrawy:1970} is larger in magnitude ($\dlong = -2\%$)
than other studies that assume a point-like pion ($\dlong =
-0.45\%$)~\cite{Bytev:2002nx,Ginsberg:1966}, the discrepancy is
believed to be calculational rather than physical.

In Sec.~\ref{sec:brem} and Sec.~\ref{sec:virt} the matrix elements
for the inner-bremsstrahlung contribution and the virtual contribution
to the radiative corrections are calculated.  The numerical methods
used to evaluate the inner-bremsstrahlung and virtual matrix elements
are discussed in Sec.~\ref{sec:numerical_method}.

\subsubsection{Inner-Bremsstrahlung Contribution}\label{sec:brem}
Previous studies of single-photon radiation in \kln decays using the
aforementioned phenomenological model are found in
Refs.~\cite{Ginsberg:pz, Ginsberg:vy, Fearing:zz, Fischbach:qw}.  In
Refs.~\cite{Ginsberg:pz, Ginsberg:vy}, the $K$--$\pi$ form factors
were taken to be constant (${\mathcal A}5$) and the final state
neutrino--photon system was assumed to be ``undetected'' (${\mathcal
A}6$).  Since the sensitivities of modern experiments are expected to
probe the form factor dependence of \kln decays, we relax ${\mathcal
A}5$.  Current experiments also measure the radiated photon from \kln
decays; therefore, we remove ${\mathcal A}6$ in our analysis.

In Refs.~\cite{Fearing:zz, Fischbach:qw}, the \klng\spac matrix
element was derived assuming a linear form factor model and using
soft-photon theorems~\cite{Low:1958sn,Burnett:1967km}.  Unlike
Refs.~\cite{Ginsberg:pz, Ginsberg:vy}, their study does not assume
that the photon is undetected; therefore, radiative photon spectra may
be generated for use in detector-acceptance studies.  In this section,
we derive the inner-bremsstrahlung matrix element for the quadratic
and the pole form factor models and we revisit the constant and linear
models.

The inner-bremsstrahlung correction to the \kln decay rate consists of
single photon radiation from the pion, the lepton, and the effective
vertex [see Fig.~\ref{fig:feynman}(g)--(i)].  The kinematics for the
inner-bremsstrahlung diagrams are described by the decay equation,
\begin{eqnarray}\label{eqn:kl3a_kin} K^0(p_K) \rightarrow
\pi^-(p_{\pi}) \; \ell^+(p_{\ell}) \; \nu_{\ell}(p_{\nu}) \; \gamma(k),
\end{eqnarray}
where $k$ is the photon four-momentum.  

Using the off-shell effective vertex,
Eq.~(\ref{eqn:effective_vertex_compact}), the contribution to the
inner-brems\-strahlung matrix element from radiation off the pion
(Fig.~\ref{fig:feynman}(g)) and the lepton (Fig.~\ref{fig:feynman}(h))
is
\begin{eqnarray}\label{eqn:brem_me_v1}
{\mathcal M}^{(g)+(h)} & = & e \sqrt{2} G_F V_{us} f_+(0)~\bar{u}(p_{\nu})P_R
\left\{ \left[ 2\sla{p}_K\hat{f}_+(t_1) + m_{\ell}\hat{f}_2(t_1)\right]
\left[\frac{\epsilon^*(k)\cdot p_{\pi}}{k\cdot p_{\pi}}\right]
\right. \nonumber \\
- \left[ 2\sla{p}_K\hat{f}_+(t_2) \right. & + & 
 \left. m_{\ell}\hat{f}_2(t_2)\right] \left.
  \left[ \frac{\epsilon^{*}(k)\cdot p_{\ell}}{k\cdot p_{\ell}}
 + \frac{\sla{k}\sla{\epsilon}^*(k)}{2k\cdot p_{\ell}} \right]
 + \hat{f}_2(t_2) \, \sla{\epsilon}^*(k) \right\} v(p_{\ell}),
\end{eqnarray}
where $\epsilon^{\mu}(k)$ is the photon polarization vector,
$m_{\ell}$ is the mass of the charged lepton, and $\hat{f}_2(t) =
\hat{f}_+(t)-\hat{f}_-(t)$.  In Eq.~(\ref{eqn:brem_me_v1}), the form
factors are evaluated for two different values of the momentum
transfer to the $W^{+}$ gauge boson: $t_1$ and $t_2$.  When the photon
is radiated from the pion, the momentum transfer is $t_1 =
(p_K-p_{\pi}-k)^2 = (p_{\ell}+p_{\nu})^2$ and when the photon is
radiated from the charged lepton the momentum transfer is
$t_2=(p_K-p_{\pi})^2$.

The last component of the inner-bremsstrahlung matrix element is from
photon radiation off the effective vertex [see
Fig.~\ref{fig:feynman}(i)].  The expression for the ``\kplng'' vertex
is obtained by requiring the matrix element to be gauge invariant.  In
Ref.~\cite{Ginsberg:pz,Ginsberg:vy} the \kplng\spac vertex is obtained
by applying the principle of ``minimal coupling'' to the
weak-interaction Lagrangian, while Refs.~\cite{Fearing:zz,
Fischbach:qw} uses the Ward identity \cite{Ward:xp} to determine
the \kplng\spac vertex.  We illustrate both of these methods and we
use the latter to determine the forms of the \kplng\spac vertex for
non-constant form factors.

Assuming the form factors are constant, the weak-interaction
Lagrangian is
\begin{equation}\label{eqn:Lagrangian}
{\mathcal L}_W = \sqrt{2} G_F V_{us}~\bar{\psi_{\nu}}
 \gamma^{\mu} P_L \psi_{\ell}~ \left[ (f_+ + f_-) \phi_{\pi} \partial_{\mu}
 \phi_K + (f_+ - f_-) \phi_{K} \partial_{\mu} \phi_{\pi} \right],
\end{equation}
where $\psi$ and $\phi$ denote the fermion and complex scalar fields,
respectively.  Replacing the partial derivative with the covariant derivative (minimal coupling),
\begin{eqnarray}\label{eqn:min_coupling}
\partial_{\mu} \rightarrow D_{\mu} = \partial_{\mu} - ieQA_{\mu},
\end{eqnarray}
where $A_{\mu}$ is the photon field, one obtains the \kplng\spac
effective vertex,
\begin{eqnarray}\label{eqn:vertex_rad_const}
V^{K\pi\ell\nu\gamma,\; \mu}_{\rm cons}
 & = & -i \sqrt{8\pi\alpha} \; G_F V_{us} P_R \; f_2 \gamma^{\mu}.
\end{eqnarray}
Combining all three inner-bremsstrahlung graphs, the matrix element
for the constant form factors is
\begin{equation}\label{eqn:brem_me_cons}
{\mathcal M}_{\rm cons}^{\rm brem}
 =  e \sqrt{2} G_F V_{us} f_+(0) \; \bar{u}(p_{\nu})P_R
 \left[ 2\sla{p}_K + m_{\ell} \right] 
 \left[ \frac{\epsilon^*(k)\cdot p_{\pi}}{k\cdot p_{\pi}} - \frac{\epsilon^*(k)
 \cdot p_{\ell}}{k\cdot p_{\ell}} - \frac{\sla{k}\sla{\epsilon}^*(k)}
 {2k\cdot p_{\ell}} \right] v(p_{\ell}).
\end{equation}
Since $\hat{f}_{+,\,2}$ are normalized and independent of $t$ they
have been set to $1$.  Taking $\epsilon^{\mu}(k) \rightarrow k^{\mu}$
in Eq.~(\ref{eqn:brem_me_cons}), one observes that ${\mathcal
M}_{\rm cons}^{\rm brem} \rightarrow 0$ and therefore the Ward identity is
satisfied.

Extraction of the \kplng\spac vertex through minimal coupling works
well for the constant form factor model.  When form factors are
no longer constant and when the matrix element depends on two
different values of $t$ ({\it e.g.}  $t_1$ and $t_2$), this method is
insufficient.  For this scenario, we determine the \kplng\spac vertex
by requiring the inner-bremsstrahlung matrix element to be zero when
the photon polarization vector $\epsilon^{\mu}(k)$ is replaced by the
photon four-momentum (Ward identity).  This method does not uniquely
determine the \kplng\spac vertex; the vertex expression may differ by
{\it structure-dependent terms}.  For a discussion of
structure-dependent terms we refer the reader to
Ref.~\cite{Fearing:zz, Fischbach:qw}.

Using this methodology, the \kplng\spac vertex for the linear form
factor model is
\begin{equation}\label{eqn:kplng_line}
V^{K\pi\ell\nu\gamma,~\mu}_{\rm lin}
 = -i\sqrt{8\pi\alpha} \; G_F V_{us} f_+(0) P_R \left\{
 \hat{f}_2(t_2)\, \gamma^{\mu} - 2\hat{f}^{\prime}_+(0)
 \left[2\sla{p}_K-\sla{p}_{\ell}\right] (p_K-p_{\pi})^{\mu} \right\},
\end{equation}
where $\hat{f}^{\prime}_+(0)$ ($=\lambda_+/m_{\pi}^2$) is the slope of
the normalized form factor.  The vertex interaction for the quadratic
model is
\begin{equation}\label{eqn:kplng_quad}
\renewcommand\arraystretch{1.5}
\begin{array}{ll}
\multicolumn{2}{l}{V^{K\pi\ell\nu\gamma,\, \mu}_{\rm quad}
 = -i\sqrt{8\pi\alpha} \; G_F V_{us} f_+(0) P_R \left\{
 \hat{f}_2(t_2)\; \gamma^{\mu} \right. } \\
 \hspace{8mm} & - 2 \left[2\sla{p}_K-\sla{p}_{\ell}\right]
 \left[ \hat{f}^{\prime}_+(0) + \frac{1}{2}(t_1+t_2)
 \hat{f}^{\prime\prime}_+(0) \right] (p_K-p_{\pi})^{\mu} \\ 
 & - \left. \sla{p}_{\ell} (m_K^2-m_{\pi}^2) \left[ \hat{f}^{\prime\prime}_0(0)
 - \hat{f}^{\prime\prime}_+(0) \right] (p_K-p_{\pi})^{\mu} \right\}, \nonumber
\end{array}
\end{equation}
where $\hat{f}^{\prime}_{+,\, 0}(0)$ ($=\lambda^{\prime}_{+,\,
0}/m_{\pi}^2$) and $\hat{f}^{\prime\prime}_{+,\, 0}(0)$
($=\lambda^{\prime\prime}_{+,\, 0}/m_{\pi}^4$). The vertex interaction
for the pole model is
\begin{equation}\label{eqn:kplng_pole}
\renewcommand\arraystretch{1.5}
\begin{array}{ll}
\multicolumn{2}{l}{V^{K\pi\ell\nu\gamma,~\mu}_{\rm pole} = -i\sqrt{8\pi\alpha}
 ~G_F V_{us} f_+(0) P_R \left\{ \hat{f}_2(t_2)\; \gamma^{\mu} \right. } \\
 \hspace{8mm} & - 2\left[2\sla{p}_K-\sla{p}_{\ell}\right] D_+(t_1,t_2)
 (p_K-p_{\pi})^{\mu} \\
              & - 2\sla{p}_{\ell} (m_K^2-m_{\pi}^2)
 \left(\frac{\mplussq-\mzerosq}{\mplussq\mzerosq}\right)(p_K-p_{\pi})^{\mu} \\
              & \;\;\;\; \left. \times \left[ f_0(t_1) D_+(t_1,t_2) + f_+(t_2)
 D_0(t_1,t_2) \right] \right\},
\end{array}
\end{equation}
where $m_+$ and $m_0$ are the poles of the $K^*$ mesons and
\begin{eqnarray}
D_{+,\, 0}(t_1,t_2) & = & m^{-2}_{+,\, 0} \, f_{+,\, 0}(t_1)f_{+,\, 0}(t_2).
\end{eqnarray}

Combining the \kplng\spac contribution with
Eq.~(\ref{eqn:brem_me_v1}), the inner-bremstrahlung matrix element is
\begin{eqnarray}\label{eqn:brem_me_v2}
{\mathcal M}^{\rm brem} &=& e \sqrt{2} G_F V_{us} f_+(0) \, \bar{u}(p_{\nu})P_R
 \left\{ \left[ 2\sla{p}_K\hat{f}_+(t_1) + m_{\ell}\hat{f}_2(t_1)\right]
 \left[\frac{\epsilon^*(k)\cdot p_{\pi}}{k\cdot p_{\pi}}\right] \right.
 \nonumber \\
 & - & \left. \left[ 2\sla{p}_K\hat{f}_+(t_2) + m_{\ell}\hat{f}_2(t_2)\right]
 \left[ \frac{\epsilon^{*}(k)\cdot p_{\ell}}{k\cdot p_{\ell}} +
 \frac{\sla{k}\sla{\epsilon}^*(k)}{2k\cdot p_{\ell}} \right]
 + {\mathcal X} \right\} v(p_{\ell}),
\end{eqnarray}
where ${\mathcal X}$ is non-zero for non-constant form factor models.
The three following equations define ${\mathcal X}$ for the linear,
the quadratic, and the pole model:
\begin{equation}\label{eqn:brem_xvals}
\renewcommand\arraystretch{2}
\begin{array}{lcl}
{\mathcal X}_{\rm lin}
 & = & 2\xi \hat{f}^{\prime}_+(0) \left[2\sla{p}_K-\sla{p}_{\ell}\right] \\ 
 {\mathcal X}_{\rm quad} & = & 2\xi \left[2\sla{p}_K-\sla{p}_{\ell}\right]
 \left[\hat{f}^{\prime}_+(0) + \frac{1}{2}(t_1+t_2) \hat{f}^{\prime\prime}_+(0)
 \right] - \xi \sla{p}_{\ell} (m_K^2-m_{\pi}^2)
 \left[ \hat{f}^{\prime\prime}_+(0) - \hat{f}^{\prime\prime}_0(0) \right] \\
 {\mathcal X}_{\rm pole}
 & = &  2\xi \left[2\sla{p}_K-\sla{p}_{\ell}\right] D_+(t_1,t_2) \\
 &  & + 2\xi \sla{p}_{\ell} (m_K^2-m_{\pi}^2) \left(\frac{\mplussq-\mzerosq}
 {\mplussq\mzerosq}\right) \left[ f_0(t_1) D_+(t_1,t_2) + f_+(t_2) D_0(t_1,t_2)
 \right],
\end{array}
\end{equation}
where $\xi = \epsilon^*(k)\cdot(p_K-p_{\pi})$, and $m_{+}$ and $m_{0}$
are the vector and scalar $K^*$ masses of the pole form factor model.

The terms in Eq.~(\ref{eqn:brem_xvals}) are necessary to maintain the
gauge invariance of the matrix element.  In our numerical evaluation
of the inner-bremsstrahlung matrix element [see
Sec.~\ref{sec:numerical_method}], we introduce a small photon mass
$\lambda$ to regulate the infrared divergence.  If the contribution
from Fig.~\ref{fig:feynman}(i) to the matrix element was not included,
then the matrix element would grow like $\sim E_{\gamma}/\lambda$.
This behavior arises from the longitudinal component of the photon
polarization vector in the term $f_2(t_2)\sla{\epsilon}(k)$ and also
from the the fact that $t_1 \neq t_2$.  Including the effect of
Fig.~\ref{fig:feynman}(i) in the matrix element, the Ward identity is
maintained and the matrix element has the correct behavior as the
photon mass, $\lambda$, goes to zero.

\subsubsection{Virtual Contribution}\label{sec:virt}
The first-order virtual corrections consist of three types of
diagrams: self-energy corrections (Fig.~\ref{fig:feynman}(b)--(c)),
inter-particle photon exchange (Fig.~\ref{fig:feynman}(a)), and
effective vertex interactions (Fig.~\ref{fig:feynman}(d)--(f)).
Previous studies~\cite{Ginsberg:pz, Ginsberg:vy} calculate the virtual
corrections assuming that the form factors are constant.  In this
section, we recalculate the virtual matrix element assuming constant
form factors and we investigate two methods to calculate the virtual
contribution for non-constant form factors.

Self-energy corrections to the pion and to the charged lepton are
obtained by wavefunction renormalization of the weak Lagrangian [see
Eq.~(\ref{eqn:Lagrangian})].  The wavefunction renormalizations for
the pion and the charged lepton are
\begin{eqnarray}\label{eqn:counterterms}
\delta Z_{\ell} & = & \frac{\alpha}{4\pi} \left[ 2B_1(\mlsq;\mlsq,\lambda^2)
 +4\mlsq B^{\prime}_1(\mlsq;\mlsq,\lambda^2)
 +8\mlsq B^{\prime}_0(\mlsq;\mlsq,\lambda^2) \right] \\
\delta Z_{\pi}  & = & \frac{\alpha}{4\pi}
 \left[ 1+ 2B_0(\mpisq;\mpisq,\lambda^2)
 + (4\mpisq-\lambda^2) B^{\prime}_0(\mpisq;\mpisq,\lambda^2) \right],
\end{eqnarray}
where $B(\cdot)$ and $B^{\prime}(\cdot)$ are the two-point loop
integral~\cite{'tHooft:1978xw} and its derivative, \\
$B^{\prime}(p_1^2,m_0^2,m_1^2) = \partial
B(p_1^2,m_0^2,m_1^2)/\partial p_1^2$.  The loop integrals are
calculated using a cutoff regularization and are defined in
Appendix~\ref{appendix:a}.  The self-energy contribution to the
virtual matrix element is
\begin{eqnarray}\label{eqn:self_energy}
{\mathcal M}^{(b)+(c)} & = & \frac{1}{2}(\delta Z_{\ell}+\delta Z_{\pi}) 
                             {\mathcal M}^{B},
\end{eqnarray}
where ${\mathcal M}^{B}$ is the lowest-order \kln matrix element
[see Eq.~(\ref{eqn:born_me})].

The inter-particle photon exchange diagram is calculated using
Eq.~(\ref{eqn:effective_vertex_compact}) for the \kpln\spac vertex.
Since the form factors are assumed constant, they may be pulled
outside of the photon four-momentum integration.  For non-constant
form factor models, the form factors will depend on the photon
four-momentum and directly modify the photonic loop integration.

For the constant form factor model, the vertex correction consists of
the diagrams (d) and (e) of Fig.~\ref{fig:feynman}.  Since the minimal
coupling anzatz does not require a \kplngg\spac vertex,
Fig.~\ref{fig:feynman}(f) is neglected in the virtual correction
calculation.  The \kplng\spac vertex from Eq.~(\ref{eqn:brem_me_cons})
is used to calculate the effective vertex interaction graphs.  As with
the evaluation of the inter-particle photon exchange diagram, the
constant form factor does not modify the integral over the photon
four-momenta.

Combining the contributions from each of the virtual diagrams, the
virtual matrix element for a constant form factor model is
\begin{equation}\label{eqn:virt_me}
{\mathcal M}^{\rm virt} = \frac{\alpha}{4\pi} \sqrt{2} G_F V_{us} f_+(0)~
\bar{u}(p_{\nu})P_R \left[ A \sla{p}_K - B \sla{p}_{\ell}\ \right] v(p_{\ell}),
\end{equation}
where $A$ and $B$ may be written in terms of photon loop integrals.
$A$ and $B$ are given by the equations
\begin{equation}\label{eqn:virt_a}
\renewcommand\arraystretch{1.5}
\begin{array}{ll}
\multicolumn{2}{l}{A = \;\; \hat{f}_+ \left[ 1+ 4B_0(\mpisq;\mpisq,\lambda^2) +2B_1(\mlsq;\mlsq,\lambda^2) \right.} \\
\hspace{10mm} & \left. + (4\mpisq -\lambda^2) B^{\prime}_0(\mpisq;\mpisq,\lambda^2) \right. \\
              & \left. + 8\mlsq B^{\prime}_0(\mlsq;\mlsq,\lambda^2) + 4\mlsq B_1^{\prime}(\mlsq;\mlsq,\lambda^2) \right. \\
              & \left. - 8x_{\pi\ell}\, C_0 + 4\mpisq C_1 - 4\left( 2x_{\pi\ell} + \mlsq \right) C_2 \right] \\
\multicolumn{2}{l}{\hspace{7mm} + \hat{f}_2 \left[ 2B_0(s_{\pi\ell};\mpisq,\mlsq) - B_0(\mpisq;\mpisq,\lambda^2) \right.} \\ 
              & \left. +4B_0(\mlsq;\mlsq,\lambda^2) + 2B_1(\mlsq;\mlsq;\lambda^2) \right. \\
              & \left. +B_1(\mpisq;\mpisq,\lambda^2) +B_1(\mpisq;\lambda^2,\mpisq) \right. \\
              & \left. +2\lambda^2 C_0 - 4x_{\pi\ell}\, C_1 + 2\mlsq C_2 \right] \\
\end{array}
\end{equation}
and
\begin{equation}\label{eqn:virt_b}
\renewcommand\arraystretch{1.5}
\begin{array}{ll}
\multicolumn{2}{l}{B = \;\; \hat{f}_+ \left[ -4 s_{\pi\ell}\, C_2 \right] } \\
\multicolumn{2}{l}{\hspace{7mm} + \hat{f}_2 \left[ \frac{1}{2} + 2B_0(s_{\pi\ell};\mpisq,\mlsq) + B_0(\mpisq;\lambda^2,\mpisq) \right.} \\
\hspace{10mm} & \left. -4B_0(\mlsq;\mlsq,\lambda^2) + B_1(\mpisq;\mpisq,\lambda^2) \right. \\
              & \left. +B_1(\mpisq;\lambda^2,\mpisq) - B_1(\mlsq;\mlsq,\lambda^2) \right. \\
              & \left. +4\mlsq B^{\prime}_0(\mlsq;\mlsq,\lambda^2) \right. \\ 
              & \left. +\frac{1}{2}(4\mpisq - \lambda^2) B^{\prime}_0(\mpisq;\mpisq,\lambda^2) \right. \\
              & \left. +2\mlsq B^{\prime}_1(\mlsq;\mlsq,\lambda^2) \right. \\
              & \left. +(2\lambda^2-4x_{\pi\ell})C_0 + (2\mpisq-4x_{\pi\ell})C_1 \right. \\
              & \left. +( 2\mlsq - 4x_{\pi\ell} ) C_2 \right] \\
\end{array}
\end{equation}
where $s_{ij}=(p_i+p_j)^2$, $x_{ij} = p_i\cdot p_j$, and $C_i =
C_i(\mpisq,s_{\pi\ell},\mlsq;\lambda^2,\mpisq,\mlsq)$.  

In Eq.~(\ref{eqn:virt_a})--(\ref{eqn:virt_b}), a small photon mass
$\lambda$ is introduced to regulate the infrared (IR) divergence.
Since the \kpln\spac vertex is an effective interaction, the
interaction is not renormalizable and will depend on an ultraviolet
(UV) cutoff scale, $\Lambda$.  Assuming a constant form factor model,
the virtual matrix element is logarithmically divergent (${\mathcal
M}^{\rm virt} \sim \ln\Lambda^2$).  A cutoff regulator is used when
calculating the loop integrals; we select the UV cutoff such that
$\Lambda > m_{\pi},\, m_{\ell}$.  We take $\Lambda$ to be the proton
mass ($m_p$).  In Sec.~\ref{sec:results}, we discuss the sensitivity
of the radiative correction to the UV cutoff scale.

For non-constant form factors,
Eq.~(\ref{eqn:virt_a})--(\ref{eqn:virt_b}) will be modified.  In this
paper, we use two methods to study the effects of non-constant form
factors in the virtual calculation.  The first method, Method I, is an
approximate method in which the constant form factors $f_{+,\,0}$ in
Eq.~(\ref{eqn:virt_a})--(\ref{eqn:virt_b}) are replaced by
$f_{+,\,0}(t)$, where $t=(p_K - p_{\pi})^2$.  Using this method, each
\kln event is modified based on the size of the momentum transfer to
the lepton--neutrino system.

The second method, Method II, assumes the pole form factor model and
it includes the effect of the photon four-momenta on the form factor.
For instance, in the inter-particle photon exchange diagram
(Fig.~\ref{fig:feynman}(a)), the form factor depends on the momentum
transfer $t_1 = (p_K-p_{\pi}-k)^2$.  Therefore, the $k$-dependence of
the form factor will modify the photonic loop integration.
Expressions for $A$ and $B$ in the Method II are in
Appendix~\ref{appendix:b}.  For Method II, we also use cutoff
regularization; therefore, a new UV cutoff scale is selected that is
larger than the mass of the $K^*$ poles.

Results from both methods are presented in Sec.~\ref{sec:results}.

\section{Numerical Methods}\label{sec:numerical_method}
To determine a numerical value for the long-distance radiative
correction, \dlongl, we integrate the squared, summed
inner-bremsstrahlung and virtual matrix elements over their respective
final-state phase spaces.  To aid in this evaluation, we developed a
program named {\tt KLOR} (Kaon Leading Order Radiation).  Using {\tt
KLOR}, we numerically evaluate the inner-bremsstrahlung and the
virtual matrix elements for both constant and non-constant form factor
models.  In this section, we discuss the methodology of the numerical
calculation.

Using the expressions for the inner-bremsstrahlung and the virtual
matrix elements in Sec.~\ref{sec:model_rad}, {\tt KLOR} numerically
squares each matrix element and sums each of them over polarization-
and spin- states.  The squared, summed matrix elements are integrated
over phase space by two methods: Vegas integration~\cite{vegas} and
traditional Monte Carlo integration~\cite{num_rec}.  The Vegas
integration is used to evaluate the total decay rate, while the
traditional Monte Carlo integration is used to check the Vegas
integration and to evaluate the decay rate when final-state
requirements on the photon kinematics ({\it e.g.}  angular and/or
energy constraints) are imposed.  An optimized, phase-space sampling
technique is used to efficiently generate unweighted event samples.

Since our calculation of the decay rate is numerical, proper
regulation of the IR divergence is necessary to obtain finite results.
The IR divergence is regulated by introducing a small photon mass
$\lambda$.  For the inner-bremsstrahlung calculation, the photon mass
limits the kinematically accessible four-body phase space.  Since the
photon is massive, one needs to sum over the longitudinal polarization
in addition to the two transverse polarizations.  In the virtual
calculation, the photon mass modifies the loop integration over the
photon momentum.  Tests of the numerical stability of our IR
regularization scheme show that {\tt KLOR}'s results are numerically
stable for photon masses from $\sim 10^{-5}$ to $10^{-14}$ GeV/c$^2$.

The phenomenological \kpln\spac interaction is effective
(non-renormalizable); therefore, this model describes the interaction
up to a UV cutoff scale, $\Lambda$.  We take the UV cutoff, $\Lambda$,
to be the {\it low-energy cutoff} described in
Sec.~\ref{sec:model_rad}; recall that the low-energy cutoff is the
boundary between the long- and short- distance corrections.  The UV
cutoff is used to regulate the loop integration -- {\it cutoff
regularization}.  In particular, the photon energy is integrated up to
the UV scale $\Lambda$.

Loop integrals in the virtual matrix element are computed using the
numerical package {\tt Looptools}~\cite{looptools,ff}.  {\tt
Looptools} calculates the value of loop integrals using analytic
expressions derived for dimensional regularization.  As long as the
cutoff scale is taken to be large; evaluation of the loop integrals in
the cutoff regularization can be related to the corresponding
expressions in dimensional regularization with minor modifications.

In Appendix~\ref{appendix:c}, we compare results from {\tt KLOR} to
other approximate techniques.  In particular, we analyze distributions
produced by {\tt KLOR} and another program called {\tt
PHOTOS}~\cite{Barberio:1993qi}.  {\tt PHOTOS} is a Monte Carlo event
generator that can be used to approximate the radiative corrections to
the \kln decay modes.

\section{Results}\label{sec:results}
In Sec.~\ref{sec:model_rad}, we developed a phenomenological framework
to describe radiative corrections to the \kln decay modes and in
Sec.~\ref{sec:numerical_method}, we described a program, {\tt KLOR},
that was written to implement this framework.  In this section, we use
{\tt KLOR} to illustrate the structure of the corrections, to
determine the \dlongl\spac parameter, and to calculate experimentally
measurable observables of the model.  

\begin{figure}[h]
\begin{center}
\includegraphics[scale=0.6]{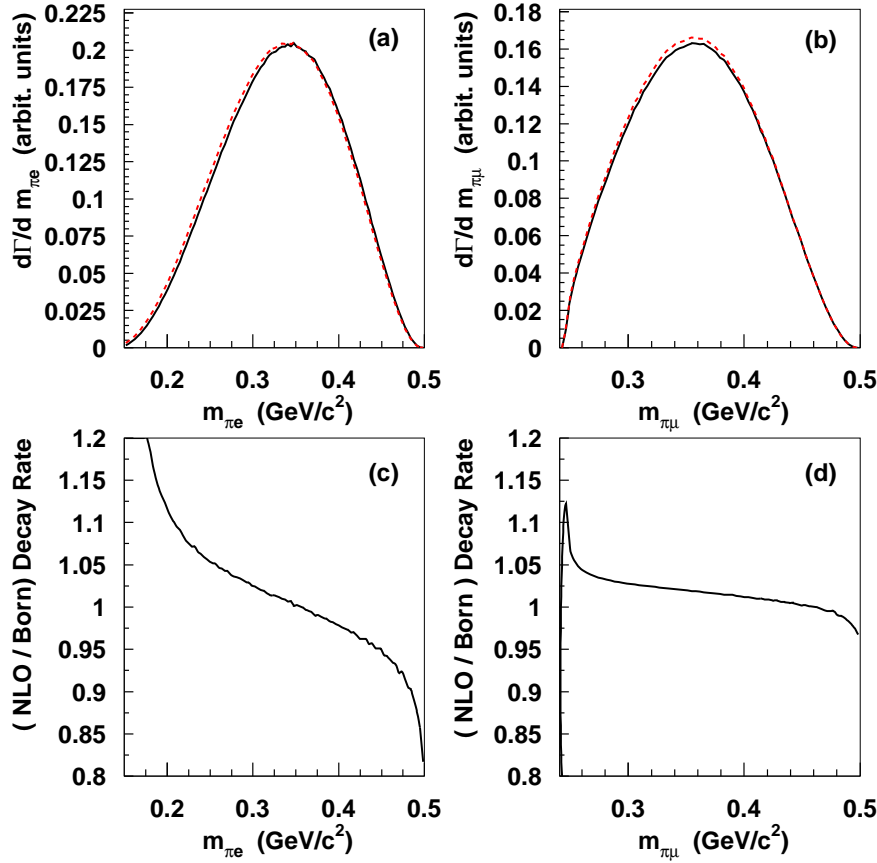}
\end{center}
\caption{\label{fig:mpil} (a) The invariant mass distribution of the
pion--electron system ($m_{e\pi}$) using the Born (solid line) and the
NLO (dashed line) matrix elements.  (b) The invariant mass
distribution of the pion--muon system ($m_{\mu\pi}$) using the Born
(solid line) and the NLO (dashed line) matrix elements.  (c) and (d)
are the ratio of the NLO and the Born distributions for the $m_{e\pi}$
and $m_{e\pi}$ distributions, respectively.}
\end{figure}

Before delving into the aforementioned topics, we would like to
illustrate the size of the radiative corrections.  One approach is to
examine the invariant mass of the pion/lepton system, denoted
$m_{\pi\ell}$.  In Fig.~\ref{fig:mpil}, we plot the Born and the
next-to-leading order (NLO) differential decay rates as a function of
$m_{\pi e}$ (Fig.~\ref{fig:mpil}(a)) and $m_{\pi\mu}$
(Fig.~\ref{fig:mpil}(b)).  In the notation of
Eq.~(\ref{eqn:tot_rate}), the Born decay rate is $\Gamma^{(0)}_{K\ell
3}$ and the NLO decay rate is $\Gamma^{(0)}_{K\ell
3}+\Gamma^{(1)}_{K\ell 3}$.

The effect of radiative corrections on the Born decay rate is
illustrated by the shifts of the NLO distribution from the Born
distribution.  In Fig.~\ref{fig:mpil}(c) and Fig.~\ref{fig:mpil}(d),
we plot the ratio of the NLO and the Born $m_{\pi\ell}$ distributions
for the \ken and \kmn modes, respectively.  In each of these plots, we
assume a pole model [see Eq.~(\ref{eqn:pole_ff})] for the hadronic
$K$--$\pi$ form factors.  The parameters of the pole model,
$m_+=\mvector$ GeV/c$^2$ and $m_0=\mzero$ GeV/c$^2$ are from a recent
measurement by the KTeV Collaboration~\cite{Alexopoulos:2004sy}.

As observed in Fig.~\ref{fig:mpil}, radiative corrections do not
uniformly modify the $m_{\pi e}$ distribution.  Radiative corrections
enhance the Born decay rate for $m_{\pi e}$ masses below $0.3$
GeV/c$^2$ and suppress the Born decay rate for $m_{\pi e}$ masses
above $0.3$ GeV/c$^2$.  A typical correction to the $m_{\pi e}$
distribution is $\sim 5\%$ but can be greater for $m_{\pi e}$ masses
near the kinematic boundaries.  Radiative corrections to the
$m_{\pi\mu}$ distribution tend to be more uniform and generally
enhance the decay rate.  A typical \kmn correction to the $m_{\pi\mu}$
distribution is $\sim 3\%$ but can be larger near the endpoints of the
$m_{\pi\mu}$ distribution.

Now that we have a basic understanding of the size of the radiative
corrections in the \ken and the \kmn modes, we investigate the
structure of the radiative corrections.  In particular, we study
radiative corrections over the \ken and the \kmn three-body Dalitz
regions.

\subsection{Dalitz Distribution of the Radiative Corrections}\label{sec:dalitz}
\begin{figure}[h]
\begin{center}
\includegraphics[scale=0.6]{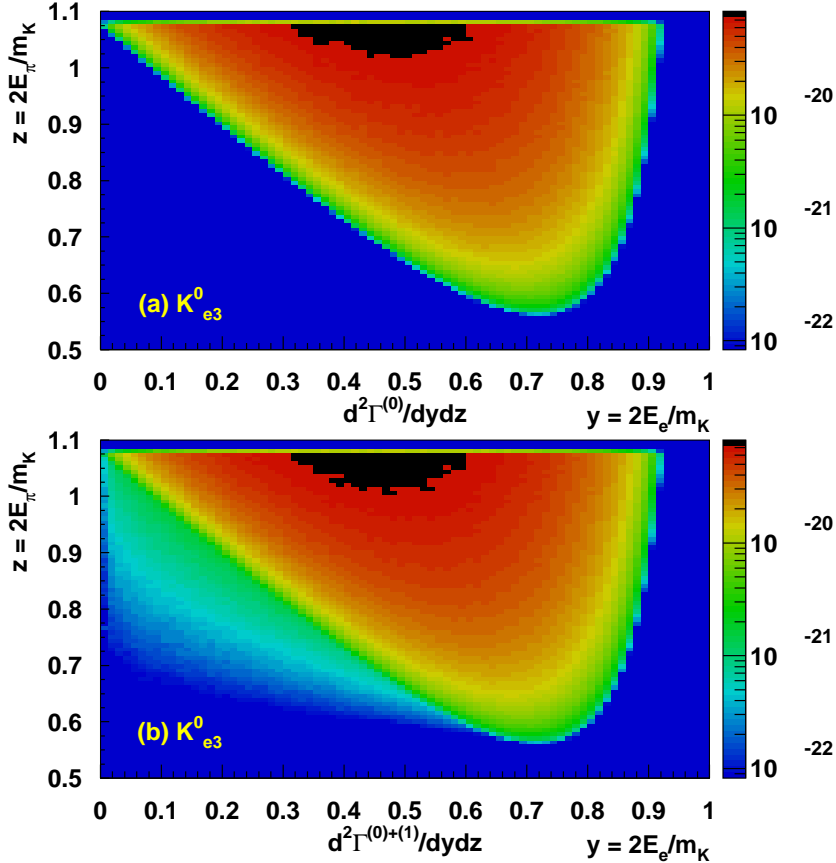}
\end{center}
\caption{\label{fig:sigma_dalitz_ke3} (a) Dalitz plot of the Born
decay rate for the \ken mode.  (b) Dalitz plot for the \ken decay rate
including radiative corrections.  Plot (b) uses the pole form factor
model ($m_+=\mvector$ GeV/c$^2$, $m_0=\mzero$
GeV/c$^2$)~\cite{Alexopoulos:2004sy} and Method I to calculate the virtual
diagrams.  The cutoff scale $\Lambda = m_p$ is used.}
\end{figure}

\begin{figure}[h]
\begin{center}
\includegraphics[scale=0.6]{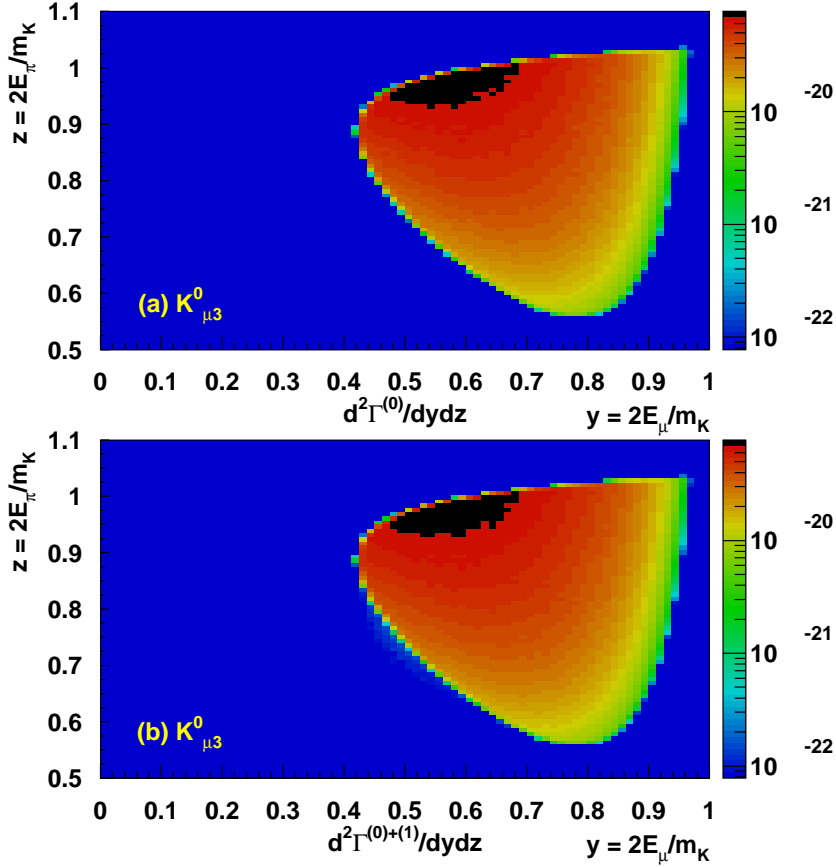}
\end{center}
\caption{\label{fig:sigma_dalitz_km3} (a) Dalitz plot of the Born
decay rate for the \kmn mode.  (b) Dalitz plot for the \kmn decay rate
including the radiative corrections.  Plot (b) uses the
pole form factor model ($m_+=\mvector$ GeV/c$^2$, $m_0=\mzero$
GeV/c$^2$)~\cite{Alexopoulos:2004sy} and Method I to calculate the virtual
diagrams.  The cutoff scale $\Lambda = m_p$ is used.}
\end{figure}

In Fig.~\ref{fig:sigma_dalitz_ke3}(a) and
Fig.~\ref{fig:sigma_dalitz_km3}(a), we plot the lowest-order decay
rates for the \ken and the \kmn decay modes over their respective
three-body Dalitz regions.  These plots illustrate the dependence of
the Born decay rate on the reduced energies
\begin{eqnarray}
y=\frac{2E_{\pi}}{m_{K}} & , & z=\frac{2E_{\ell}}{m_{K}},
\end{eqnarray}
where $E_{\pi}$ ($E_{\ell}$) is the pion (charged lepton) energy in
the kaon center of mass.  

When radiative effects are included, the Born Dalitz distributions are
modified.  Using Method I [see Sec.~\ref{sec:virt}], the NLO decay
rate for the \ken and \kmn modes are plotted in
Fig.~\ref{fig:sigma_dalitz_ke3}(b) and
Fig.~\ref{fig:sigma_dalitz_km3}(b), respectively.  In general,
radiative corrections modify the Born decay rate both inside and
outside the Dalitz region.  Inside the Dalitz region, both virtual and
inner-bremsstrahlung diagrams contribute to the NLO decay rate; while
outside the Dalitz region, only inner-bremsstrahlung diagrams
contribute.  The inner-bremsstrahlung contribution to the \kln decay
rate outside the Dalitz region is more pronounced in the \ken decay
mode than in the \kmn decay mode since there is a higher probability
of radiation from the electron.

The long-distance radiative correction to each point in the three-body
Dalitz plane, denoted $\dlongl(y,z)$, may be obtained by taking the
ratio of the NLO and the Born distributions and then subtracting
unity.  $\dlongl(y,z)$ is interesting since it describes the structure
of the radiative correction; for a given ($E_{\pi}$, $E_{\ell}$),
$\dlongl(y,z)$ characterizes how the Born distribution is modified.

\begin{figure}[h]
\begin{center}
\includegraphics[scale=0.6]{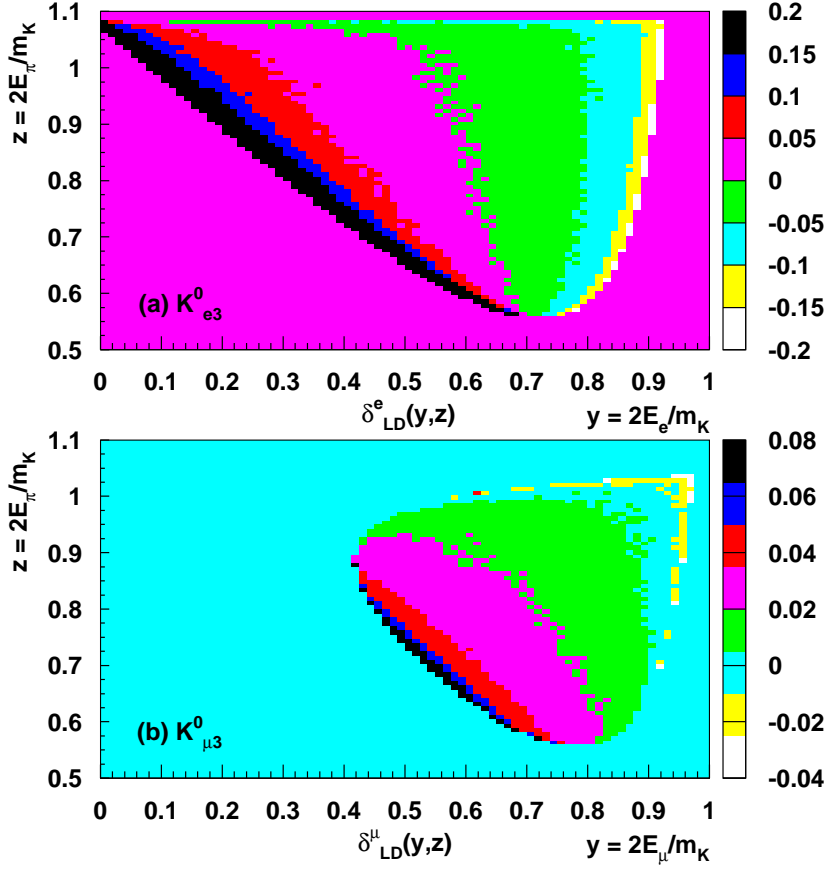}
\end{center}
\caption{\label{fig:delta} Long-distance radiative correction,
\dlongl, to \ken (a) and the \kmn (b) Dalitz plots.  Both plots use
the pole form factor model ($m_+=\mvector$ MeV/c$^2$, $m_0=\mzero$
MeV/c$^2$)~\cite{Alexopoulos:2004sy} and Method I to calculate the virtual
diagrams.  The UV cutoff ($\Lambda$) for the long-distance calculation
is taken to be the proton mass.}
\end{figure}

In Fig.~\ref{fig:delta}, we plot $\dlongl(y,z)$ using Method I for
both the \ken and the \kmn modes.  Radiative corrections to the \ken
Dalitz region can be significant, ranging from $-15$\% to $15$\%.
Radiative corrections to the \kmn Dalitz region are smaller than for
the \ken mode; they range from $-2$\% to $8$\%.  Since there is no
contribution to the Born decay rate from outside the Dalitz region,
$\dlongl(y,z)$ is infinite in this region.  In Fig.~\ref{fig:delta},
we have set $\dlongl(y,z)$ outside the Dalitz region to zero.
Depending on the technique used to measure the \kln decay rate, the
contribution from outside the Dalitz region may need to be included in
the \dlongl\spac calculation.  In Sec.~\ref{sec:delta}, we address
this and other issues affecting the \dlongl\spac calculation.

Before we calculate \dlongl, we compare the two methods, Method I and
Method II, proposed in Sec.~\ref{sec:virt}.  Recall that Method I and
Method II are two ways to introduce non-constant form factors into the
virtual matrix element calculation.  We compare these methods by
generating radiative distributions and examining the differences
between them.

\begin{figure}[h]
\begin{center}
\includegraphics[scale=0.61]{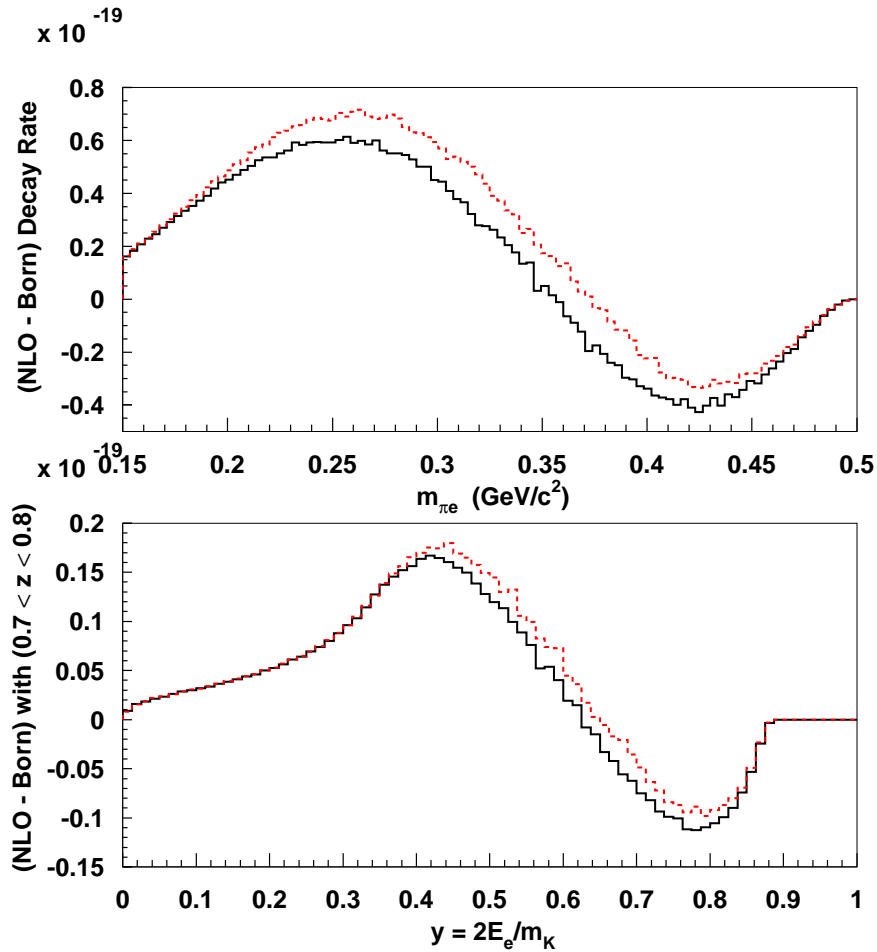}
\end{center}
\caption{\label{fig:method_compare} (a) The $m_{\pi e}$ distribution
of the difference between the NLO and the Born \ken decay rates.  Method
I (solid) and Method II (dashed) are presented.  (b) The $y$
distribution of the difference between the NLO and the Born \ken decay
rates in the $0.7<z<0.8$ band of the \ken Dalitz plane.  Method I
(solid) and Method II (dashed) are presented.}
\end{figure}

In Fig.~\ref{fig:method_compare}(a), we plot the difference between
the NLO and Born decay rate as a function of $m_{\pi e}$.  The solid
(dashed) line is produced using Method I (Method II).  The $m_{\pi e}$
distribution using Method II is shifted higher than that produced by
Method I; however, the size and structure of the distributions are
similar.

In Fig.~\ref{fig:method_compare}(b), we plot the $y$ ($=2E_e/m_K$)
distribution for the difference between the NLO and Born decay rates
in the $0.7<z<0.8$ ``band'' of the Dalitz plane.  As with the $m_{\pi
e}$ distribution, the radiative correction from the Method I and
Method II have very similar sizes and shapes.  Inspection of other
bands in the Dalitz plane lead to similar findings.

Since the radiative correction distributions produced by Method I and
Method II produce commensurate spectra, one expects that the
distribution of \kln events produced by Method I and Method II to be
identical for current experimental precision.  The distributions from
Method II are in general more positive than those of Method I;
therefore, we expect that the total radiative correction for Method II
to be slightly larger.

\subsection{The \dlongl\spac Parameter}\label{sec:delta}
The total radiative correction, denoted \dtot, to the Born \kln decay
rate consists of a short- and long-distance component.  Using the
phenomenological model [see Sec.~\ref{sec:model_rad}], we calculate
\dlongl\spac and estimate its uncertainty.  Before we can determine
\dlongl, we must first specify the analysis technique used by
experimentalists to measure the \kln decay rate.

In previous studies~\cite{Ginsberg:pz, Ginsberg:vy, Cirigliano}, the
authors assume that the experimental apparatus only measures the
charged lepton and the pion; the neutrino and (if present) the photon
are ``invisible.''  The observed pion and charged lepton momenta are
then fit to three-body kinematics with zero missing mass.  If a \kln
event produces a `$\ell\pi$' signature that lies outside the
three-body Dalitz region, then it is excluded from the decay rate
measurement.  Note that this scenario only works for an ideal
experiment for which the kaon energy is known and where the analysis
cuts are such that events can not migrate from outside the three-body
Dalitz region.  For this experimental analysis technique, the
long-distance radiative correction parameter is the total correction
to the Born decay rate {\it inside} the three-body \kln Dalitz region.
In this scenario, we denote the long-distance radiative correction
$\hat{\delta}^{\ell}_{LD}$.

We consider an experimentally motivated approach to the analysis in
which the experiment is sensitive to and detects radiated photons in
the final state.  Therefore, an analysis of the \kln decay rate will
include radiated photons.  Moreover, we also assume the experimental
analysis includes \kln signatures with measured pion and charged
lepton energies outside the three-body Dalitz region.  In this
analysis, \dlongl\spac includes contributions from both inside and
outside the three-body Dalitz plot [see
Fig.~\ref{fig:sigma_dalitz_ke3}].  Note that since we perform a
numerical analysis [see Sec.~\ref{sec:numerical_method}], one can
compute \hdlongl, which excludes events from outside the three-body
Dalitz region.

In addition to the experimental technique, \dlongl\spac will depend on
the cutoff scale $\Lambda$.  As described in Sec.~\ref{sec:model_rad},
both the short- and long-distance radiative corrections depend on the
scale $\Lambda$ which denotes the ``boundary'' between the short- and
long-distance loop corrections.  Below $\Lambda$, radiative
corrections to the \kln decay rate are described by the
phenomenological model; while, above $\Lambda$, radiative corrections
are described by the Standard Model.  In our analysis, we associate
uncertainty in the total radiative correction (\dtot) due to the
cutoff scale choice with uncertainty in the long-distance radiative
correction.

The choice of the cutoff scale is somewhat arbitrary and typically
depends on the ``physics'' and the calculational limitations of the
long-distance model.  Since we use the cutoff regulator to compute the
loop integrals, we take $\Lambda$ to be larger than the masses
appearing in the loop integrals.  In Method I and Method II, we take
$\Lambda$ to be $m_p$ and $3$ GeV/c$^2$, respectively.  Though the
assumption that the long-distance model describes physics at the
cutoff scale $\Lambda=3$ GeV/c$^2$ is likely optimistic, integration
of the loop integrals, which depend on the $K^*$ poles, requires a
higher cutoff scale.

In Table~\ref{tab:delta_ele} and Table~\ref{tab:delta_mu}, we compute
the short- and long-distance radiative corrections for various cutoff
scales ranging from $m_{\rho}$ ($=0.7711$ GeV/c$^2$) to $3$ GeV/c$^2$.
For Method I, changing the cutoff scale from $m_{\rho}$ to $3$
GeV/c$^2$ results in changes to \dlonge\spac, \dlongm\spac, and to
\dshort\spac of $\sim 0.7\%$; however, \dtote\spac ($=\dshort +
\dlonge$) remains stable to within $\sim 0.2\%$.  Therefore, a
reasonable uncertainty in \dlonge\spac due to the cutoff scale
uncertainty is $0.2\%$.  A similar inspection of the \kmn mode
indicates an uncertainty to \dlongm\spac from the cutoff scale is
$0.1\%$.

In addition to the uncertainty associated with the cutoff scale
ambiguity, there is an uncertainty in the radiative corrections due to
the method used to calculate the virtual corrections.  Using Method II,
the total radiative corrections to the \ken and the \kmn modes are
slightly larger than the corresponding values obtained via Method I.
Therefore, for \dlonge\spac and \dlongm\spac we take an additional
uncertainty of $0.1\%$ and $0.2\%$ due to the method that is used to
calculate the virtual corrections.

Therefore, after inspection of the shifts to the total radiative
corrections in Table~\ref{tab:delta_ele} and Table~\ref{tab:delta_mu},
we take the uncertainty on the long-distance radiative corrections due
to the cutoff scale and the virtual calculation method to be $0.3\%$
for both \dlonge\spac and \dlongm.
\begin{table}
\caption{\label{tab:delta_ele} Long- and
short-distance~\cite{footnote:shortdistance} radiative corrections to
the \ken decay modes.  Results are presented for different virtual
correction methods (I and II), for different form factor models, and
for various UV cutoff scales.  We use the measured values of the form
factor parameters taken from Ref.~\cite{Alexopoulos:2004sy}.}
\begin{center}
\begin{tabular}{l c c c c} \hline \hline
FF Model & $\Lambda$   & $\dshort$     & $\dlonge$     & $\dtote$       \\
         & (GeV/c$^2$) & (\%)          & (\%)          & (\%) \\ \hline
\multicolumn{5}{l}{\it Method I} \\
\hspace{2mm} Constant     & $m_{\rho}$  &  2.27         & 1.26          & 3.53           \\
\hspace{2mm} Constant     & $m_p$       &  2.17         & 1.31          & 3.48           \\
\hspace{2mm} Constant     & $2m_p$      &  1.94         & 1.55          & 3.49           \\
\hspace{2mm} Constant     & $3$         &  1.64         & 1.73          & 3.37           \\
\hspace{2mm} Linear       & $m_p$       &  2.17         & 1.32          & 3.49           \\
\hspace{2mm} Quadratic    & $m_p$       &  2.17         & 1.33          & 3.50           \\
\hspace{2mm} Pole         & $m_p$       &  2.17         & 1.33          & 3.50           \\ \hline
\multicolumn{5}{l}{\it Method II} \\
\hspace{2mm} Pole         & $m_p$       &  2.17         & 1.43          & 3.60           \\
\hspace{2mm} Pole         & $2m_p$      &  1.94         & 1.84          & 3.78           \\
\hspace{2mm} Pole         & $3$         &  1.64         & 2.11          & 3.75 
 \\
\hline \hline
\end{tabular}
\end{center}
\end{table}

\begin{table}
\caption{\label{tab:delta_mu} Long- and
short-distance~\cite{footnote:shortdistance} radiative corrections to
the \kmn decay modes.  Results are presented for different virtual
correction methods (I and II), for different form factor models, and
for various UV cutoff scales.  We use the measured values of the form
factor parameters taken from Ref.~\cite{Alexopoulos:2004sy}.}
\begin{center}
\begin{tabular}{l c c c c} \hline \hline
FF Model                  & $\Lambda$   & $\dshort$     & $\dlongm$     & $\dtotm$       \\
                          & (GeV/c$^2$) & (\%)          & (\%)          & (\%)           \\ \hline
\multicolumn{5}{l}{\it Method I} \\
\hspace{2mm} Constant     & $m_{\rho}$  &  2.27         & 1.79          & 4.06           \\
\hspace{2mm} Constant     & $m_p$       &  2.17         & 1.88          & 4.05           \\
\hspace{2mm} Constant     & $2m_p$      &  1.94         & 2.17          & 4.11           \\
\hspace{2mm} Constant     & $3$         &  1.64         & 2.35          & 3.99           \\
\hspace{2mm} Linear       & $m_p$       &  2.17         & 1.93          & 4.10           \\
\hspace{2mm} Quadratic    & $m_p$       &  2.17         & 1.93          & 4.10           \\
\hspace{2mm} Pole         & $m_p$       &  2.17         & 1.92          & 4.09           \\ \hline
\multicolumn{5}{l}{\it Method II} \\
\hspace{2mm} Pole         & $m_p$       &  2.17         & 1.96          & 4.13           \\
\hspace{2mm} Pole         & $2m_p$      &  1.94         & 2.38          & 4.32           \\
\hspace{2mm} Pole         & $3$         &  1.64         & 2.66          & 4.30
 \\ \hline \hline
\end{tabular}
\end{center}
\end{table}

\begin{figure}[h]
\begin{center}
\includegraphics[scale=0.6]{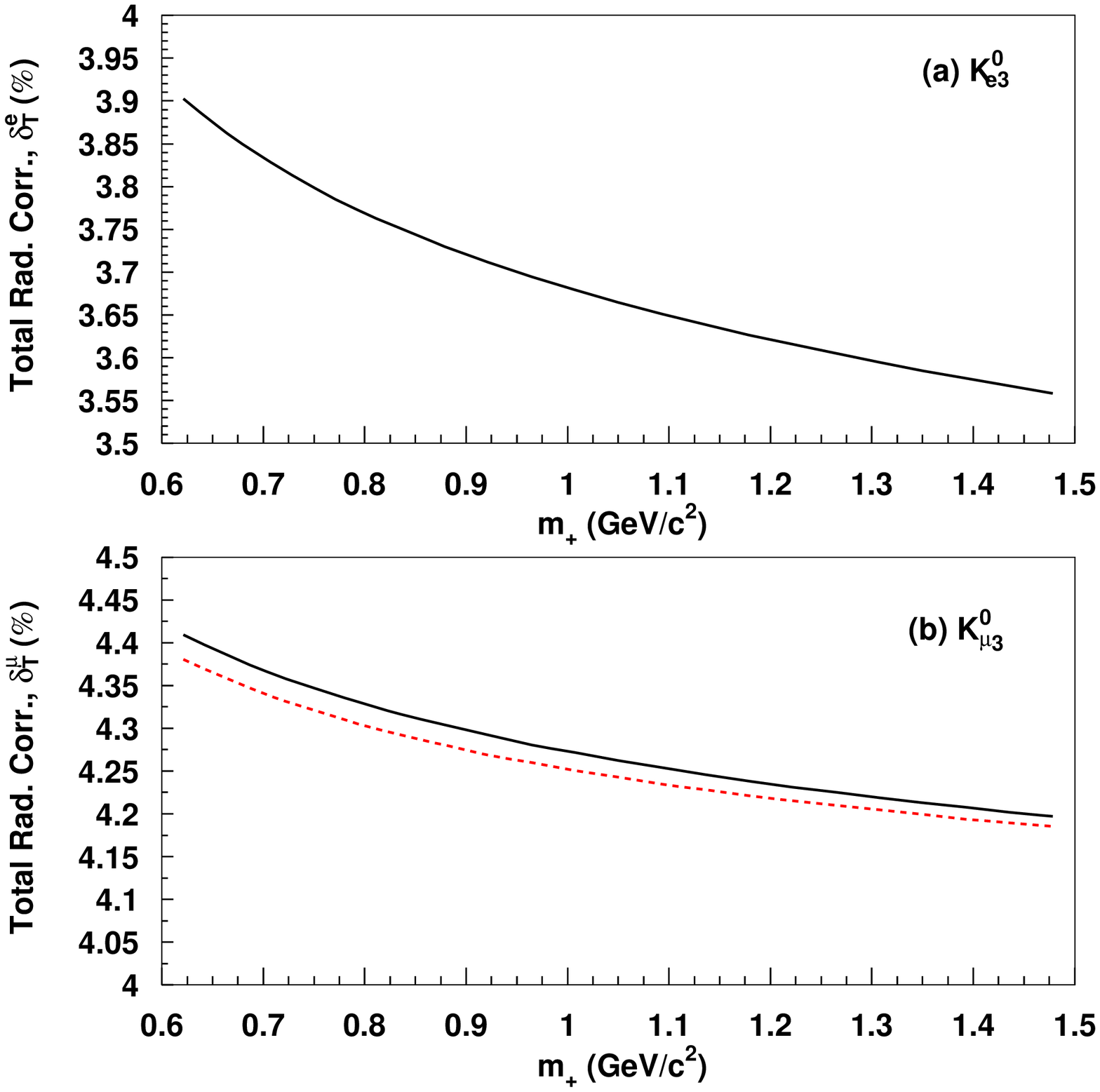}
\end{center}
\caption{\label{fig:totcor} (a) Total radiative correction, $\dtote$
($=\dlonge+\dshort)$, to the \ken decay rate as a function of the
$K^*$ pole mass $m_+$.  Curve is generated using Method II and
$m_0=\mzero$ GeV/c$^2$.  (b) Total radiative correction, $\dtotm$
($=\dlongm+\dshort)$, to the \ken decay rate as a function of the
$K^*$ pole mass $m_+$.  Curves are generated using Method II and
$m_0=\mzero$ GeV/c$^2$ (solid) and $m_0=0.9$ GeV/c$^2$ (dashed).}
\end{figure}

Using Method I and Method II [see Sec.~\ref{sec:virt}], we also study
the sensitivity of the radiative corrections to variations in the form
factor parameters.  In Fig.~\ref{fig:totcor}, we use Method II to
calculate the total radiative corrections \dtote\spac and \dtotm\spac
as a function of the pole model parameter $m_+$; the pole model
parameter $m_0$ is held constant in these plots.  In
Fig.~\ref{fig:totcor}(a), the pole model parameter $m_0=\mzero$ and in
Fig.~\ref{fig:totcor}(b), we generate the curves using $m_0=\mzero$
GeV/c$^2$ and $m_0=0.9$ GeV/c$^2$.  For the \ken decay mode,
\dlonge\spac is insensitive to $m_0$ since the $f_0(t)$ form factor is
multiplied by the square of the electron mass.

Using Method I, we find that the total radiative corrections to the
\ken and \kmn modes are insensitive to changes in the $m_+$.  For
Method I, the insensitivity of \dlonge\spac and \dlongm\spac to
changes in the form factor parameters is likely a result of the
simplistic method used to introduce $t$ dependence of the form factors
into the virtual correction, $f_{+,0} \rightarrow f_{+,0}(t)$ [see
Sec.~\ref{sec:virt}].

Using Method II, the total radiative correction shows variation with
changes in the pole model parameters.  Using this method the total
radiative correction varies $\sim 0.3\%$ for the \ken and $\sim 0.2\%$
for the \kmn modes over the $m_+$ range $0.6$--$1.5$ GeV/c$^2$.
Though Method II indicates variation of \dlonge\spac and \dlongm\spac
with changes in the form factor parameters, for reasonable changes in
the form factor parameters the variation to the radiative correction
parameters are small ($\sim 0.04\%$).  Therefore, the assumption that
the variation of the $K$--$\pi$ form factor parameters produce a
negligible effect on the radiative corrections is reasonable.

Using the phenomenological model and a cutoff scale of $\Lambda=m_p$,
we find that the long-distance radiative corrections to the \ken and
the \kmn decay modes are $1.3\%$ and $1.9\%$, respectively.  For this
model, the uncertainties on \dlonge\spac and \dlongm\spac result from
the cutoff scale/virtual calculation method ($0.3$\%) and the form
factor parameter dependence ($0.04$\%).  Therefore, we take the
uncertainty on \dlonge\spac and \dlongm\spac to be $\sim 0.3\%$.

In previous studies of the long-distance radiative corrections,
Refs.~\cite{Ginsberg:pz,Ginsberg:vy} found that $\hdlonge = 1.5\%$ and
$\hdlongm = 2.0\%$~\cite{footnote:calculation_errors}.  Recall that
\hdlongl\spac ($\ell = e$, or $\mu$) corresponds to the long-distance
radiative correction inside the \kln three-body Dalitz regions.  Since
$\hdlongl < \dlongl$, one infers that $\dlonge > 1.5\%$ and $\dlongm >
2.0\%$ from Ref.~\cite{Ginsberg:pz,Ginsberg:vy}.  Using our
calculation with a constant form factor model, $\dlonge = 1.3\%$ and
$\dlongm = 1.9\%$ [see Table~\ref{tab:delta_ele} and
Table~\ref{tab:delta_mu}].  The cause of the discrepancy between the
\dlongl\spac values is unknown; however, Ref.~\cite{Cirigliano} has
noted a formerly undocumented error in the \klc calculation of
Ref.~\cite{Ginsberg:pz,Ginsberg:vy}.  Therefore, one possible
explanation is an unknown error in the \kln calculation of
Ref.~\cite{Ginsberg:pz,Ginsberg:vy}.

A more recent study~\cite{Cirigliano} of the \ken decay rate using
\xpt\spac finds that for $\Lambda=m_{\rho}$, \hdlonge\spac is $0.6\%$.
When the cutoff scale is taken to be $m_p$, \hdlonge\spac should
increase by $\sim 0.1\%$.  In our calculation, we find that for the
\kln decay rate, the contribution to \dlonge\spac from outside the
\ken Dalitz region is $\sim +0.5\%$.  Therefore, we infer that
$\dlonge\sim 1.2\%$ from Ref.~\cite{Cirigliano}, which agrees with our
calculation.

Predictions of the phenomenological model may be tested
experimentally.  One way to test the long-distance radiative
corrections \dlonge\spac and \dlongm\spac is to test the ratio
\begin{eqnarray}\label{eqn:deltafrac}
{\mathcal F}_{LD} & = & \frac{1+\delta^{\mu}_{LD}}{1+\delta^{e}_{LD}}.
\end{eqnarray}
Using our radiative correction parameters, we find that ${\mathcal
F}_{LD} = 1.0058\pm 0.001$.  The uncertainty in ${\mathcal F}_{LD}$ is
obtained from Table~\ref{tab:delta_ele} and Table~\ref{tab:delta_mu};
note that the common cutoff scale uncertainties likely cancel in the
ratio.  Using Eq.~(\ref{eqn:tot_rate}) and assuming lepton
universality, the ratio ${\mathcal F}_{LD}$ equals
\begin{eqnarray}
\frac{\Gamma_{\kmn}}{\Gamma_{\ken}} \cdot
   \frac{I^{e}_{(0)}(\lambda_i)}{I^{\mu}_{(0)}(\lambda_i)},
\end{eqnarray}
where $\Gamma_{\kmn}/\Gamma_{\ken}$ is the ratio of the \ken and the
\kmn decay rates, $I^{e}_{(0)}(\lambda_i)/I^{\mu}_{(0)}(\lambda_i)$ is
the ratio of the lowest order \ken and \kmn phase space integrals [see
Eq.~(\ref{eqn:psintegral0})], and the $\lambda_i$ are form factor
model parameters.

Using the measured decay rates and the form factors from Ref.\
\cite{Alexopoulos:2004sw}, we find that the experimental value of
${\mathcal F}_{LD}$ is $1.0027\pm 0.0048$.  This is consistent with
our calculated value.

\subsection{Radiative Fractions}\label{sec:radfrac}
In addition to the fraction ${\mathcal F}_{LD}$ defined in
Eq.~(\ref{eqn:deltafrac}), the phenomenological model may be used to
predict other experimentally-measurable quantities.  In this section,
we use the phenomenological model to predict the fraction of radiative
events that satisfy requirements on the photon's energy and angular
distance from the charged lepton.  Radiative fractions for \kln decays
were originally proposed by Ref.~\cite{Fearing:zz} and later refined
in Refs.~\cite{Bijnens,Doncel:dy}.

The radiative fractions are defined by the equation
\begin{equation}\label{eqn:radfrac}
{\mathcal R}^0_{K\ell 3}(E^{\rm \rm min}_{\gamma},\,
 \theta^{\rm min}_{\ell\gamma}) =
 \Gamma(K^0 \rightarrow \pi^- \ell^+ \nu \gamma;\, E_{\gamma}
 \geq E^{\rm min}_{\gamma},\, \theta_{\ell\gamma}
 \geq \theta^{\rm min}_{\ell\gamma})/\Gamma_{\kln},
\end{equation}
where $E^{\rm min}_{\gamma}$ is the minimum energy of the radiated photon,
and $\theta^{\rm min}_{\ell\gamma}$ is the minimum angular separation
between the photon and the charged lepton, and $\Gamma_{\kln}$ is the
\kln decay rate including radiative corrections.  Both
$E^{min}_{\gamma}$ and $\theta^{\rm min}_{\ell\gamma}$ are quantities
defined in the kaon center of mass.

In Table~\ref{tab:radfrac}, we present the calculated radiative
fractions.  For the \ken and the \kmn decay modes we consider two
values of $E^{\rm min}_{\gamma}$: $10$ MeV and $30$ MeV.  We do not make
any requirement on the angular separation between the photon and the
muon; however, for the \ken decay mode, we also consider
$\theta^{\rm min}_{\ell\gamma}=20^{\circ}$.  The uncertainties in
Table~\ref{tab:radfrac} are composed of two components: the
uncertainty on the next-to-leading order correction and the
uncertainty on the \kln decay rates.  For the \ken and the \kmn decay
modes, we take the uncertainty on the next-order corrections to be
$1.3$\% and $1.9$\% of $\Gamma(K^0 \rightarrow \pi^- \ell^+ \nu
\gamma;\, E_{\gamma} \geq E^{\rm min}_{\gamma},\, \theta_{\ell\gamma} \geq
\theta^{\rm min}_{\ell\gamma})$, respectively.  We estimate the
uncertainties on the \kln decay rate by using the $0.3\%$ uncertainty
on the radiative correction parameters, \dlongl.
\begin{table}
\caption{\label{tab:radfrac}Radiative fraction, ${\mathcal R}^0_{K\ell
3}(E^{\rm min}_{\gamma},\, \theta^{\rm min}_{\ell\gamma})$, as defined in
Eq.~(\ref{eqn:radfrac}).  Radiative fractions are given in percent(\%)
and they are presented for both the \ken and \kmn decay modes.}
\begin{center}
\begin{tabular}{l c c c} \hline \hline
 & \multicolumn{2}{c}{${\mathcal R}^0_{Ke3}$ (\%)}
 & \multicolumn{1}{c}{${\mathcal R}^0_{K\mu 3}$ (\%)} \\ 
$E^{\rm min}_{\gamma}$ / $\theta^{\rm min}_{\ell\gamma}$  & $0^{\circ}$      &  $20^{\circ}$    & $0^{\circ}$          \\ \hline
$10$ MeV                                          &   $4.93\pm 0.06$ &  $1.89\pm 0.02$  & $0.564\pm 0.01$   \\
$30$ MeV                                          &   $2.36\pm 0.03$ &  $0.956\pm 0.01$ & $0.214\pm 0.004$
 \\ \hline \hline
\end{tabular}
\end{center}
\end{table}

\section{Conclusions}
\label{sec:conclusion}
We have studied radiative corrections to the \ken and the \kmn decay
modes using a phenomenological model for their interactions.  In our
calculation, we include the contribution to the radiative correction
from outside the three-body \kln Dalitz region.  Using the cutoff
scale $\Lambda = m_p$, we find that $\dlonge = (1.3\pm 0.3)\%$ and
$\dlongm = (1.9\pm 0.3)\%$.  The uncertainty in the long-distance
radiative corrections is dominated by their sensitivity to the cutoff
scale $\Lambda$ and virtual calculation method.  Dependence of the
radiative corrections on the variation of the hadronic $K$--$\pi$ form
factors is found to be small ($0.04\%$).  The calculated value of the
$\dlonge$ radiative correction parameter agrees with a recent
calculation~\cite{Cirigliano} using \xpt.

Combining the long-distance contribution from the phenomenological
model with the short-distance contribution from Ref.~\cite{marciano},
the total radiative corrections to the \ken and the \kmn modes are
$(3.5\pm 0.3)\%$ and $(4.1\pm 0.3)\%$.  The uncertainty in the total
radiative corrections are dominated by the uncertainty in the
long-distance component.  When measuring the $V_{us}$ CKM matrix
element, the uncertainty from radiative corrections should be $\sim
0.0003$.  This uncertainty corresponds to a small fraction of the
``external uncertainty'' ($0.0021$) quoted for the $|V_{us}|$
measurement in Ref.~\cite{Alexopoulos:2004sw}.

The program {\tt KLOR} was written to numerically evaluate the
radiative corrections and to generate Monte Carlo events.  Monte Carlo
events produced by {\tt KLOR} may be used to understand experimental
detection efficiencies.

\section*{Acknowledgments}
I would like to thank Jon Rosner, Ed Blucher, Sasha Glazov, and Rick
Kessler for useful discussions.  This work was supported in part by
the United States Department of Energy under Grant No. DE FG02
90ER40560.

\appendix

% Appendix A %%%%%%%%%%%%%%%%%%%%%%%%%%%%%%%%%%%%%%%%%%%%%%%%%%%%%%%%
\section{Loop Integrals}\label{appendix:a}
Virtual corrections to the \kln decay mode require the evaluation of
loop integrals.  Below, we define the loop integrals for one-, two-,
three-, and four- point functions~\cite{looptools,ff,'tHooft:1978xw}.
The corresponding vector loop integrals and their expansion in terms
of momentum four-vectors are also defined.  In Sec.~\ref{sec:results},
we numerically evaluate the loop integrals using the cutoff
regularization scheme.  Therefore, implicit in our definition is a
high-mass, cutoff scale, $\Lambda$.

The scalar one-point function is
\begin{equation}\label{eqn:a0}
A(m^2) = \frac{1}{i\pi} \int d^4k \frac{1}{(k^2-m^2)},
\end{equation}
where $m$ is a mass parameter.

The scalar two-point function is
\begin{equation}\label{eqn:b0}
\renewcommand\arraystretch{1.5}
\begin{array}{l}
B_0(p_1^2;m_0^2,m_1^2) = \frac{1}{i\pi} \int \frac{d^4k}{(k^2-m_0^2)((k+p_1)^2-m_1^2)},
\end{array}
\end{equation}
where $p_1^{\mu}$ is a four-momentum and $m_i$ ($i=1,2$) are mass
parameters.  The vector two-point function is
\begin{equation}\label{eqn:b1}
B_{\mu}(p_1;m_0^2,m_1^2)
 = \frac{1}{i\pi} \int \frac{d^4k\,k_{\mu}}{(k^2-m_0^2)((k+p_1)^2-m_1^2)}
 = p_{1,\mu} B_1(p_1^2;m_0^2,m_1^2),
\end{equation}
where $B_{\mu}(p_1;m_0^2,m_1^2)$ is proportional to $p_{1,\,\mu}$.

The scalar three-point function is
\begin{equation}\label{eqn:c0}
C_0(p_1^2,s_{12},p_2^2;m_0^2,m_1^2,m_2^2) = \frac{1}{i\pi} \int
\frac{d^4k}{(k^2-m_0^2)((k+p_1)^2-m_1^2)((k+p_2)^2-m_2^2)}
\end{equation}
where $s_{ij} = (p_i-p_j)^2$ is the squared difference of two
four-momenta.  The vector three-point function is
\begin{eqnarray}\label{eqn:c1}
C_{\mu}(p_1,p_2;m_0^2,m_1^2,m_2^2) & = & \frac{1}{i\pi} \int
 \frac{d^4k k_{\mu}}{(k^2-m_0^2)((k+p_1)^2-m_1^2)((k+p_2)^2-m_2^2)}
 \nonumber \\ & = & p_{1,\mu} C_1 + p_{2,\mu} C_2,
\end{eqnarray}
where $C_i = C_i(p_1^2,s_{12},p_2^2;m_0^2,m_1^2,m_2^2)$.

The scalar four-point function is 
\begin{equation}\label{eqn:d0}
\renewcommand\arraystretch{1.5}
\begin{array}{ll}
\multicolumn{2}{l}{D_0(p_1^2,s_{12},s_{23},p_3^2,p_2^2,s_{13};m_0^2,m_1^2,m_2^2,m_3^2)} \\ 
\hspace{6mm} & = \frac{1}{i\pi} \int d^4k \, [(k^2-m_0^2)((k+p_1)^2-m_1^2)]^{-1} \\
             & \hspace{12mm} \times [((k+p_2)^2-m_2^2)((k+p_3)^2-m_3^2)]^{-1}
\end{array}
\end{equation}
and the vector four-point function is 
\begin{equation}\label{eqn:d1}
\renewcommand\arraystretch{1.5}
\begin{array}{ll}
\multicolumn{2}{l}{D_{\mu}(p_1,p_2,p_3;m_0^2,m_1^2,m_2^2,m_3^2)} \\ 
\hspace{6mm} & = \frac{1}{i\pi} \int d^4k \, k_{\mu} \, [(k^2-m_0^2)((k+p_1)^2-m_1^2)]^{-1} \\
             & \hspace{12.5mm} \times [((k+p_2)^2-m_2^2)((k+p_3)^2-m_3^2)]^{-1} \\
             & = p_{1,\mu} D_1 + p_{2,\mu} D_2 + p_{3,\mu} D_3,
\end{array}
\end{equation}
where $D_i =
D_i(p_1^2,s_{12},s_{23},p_3^2,p_2^2,s_{13};m_0^2,m_1^2,m_2^2,m_3^2)$.

% Appendix B %%%%%%%%%%%%%%%%%%%%%%%%%%%%%%%%%%%%%%%%%%%%%%%%%%%%%%%%
\section{Method II Expressions}\label{appendix:b}
As in Method I, the matrix element for the virtual corrections using
Method II may be written in the following form:
\begin{equation}\label{eqn:virt_me_pole}
{\mathcal M}^{\rm virt}_{\rm pole} = \frac{\alpha}{4\pi}
 \sqrt{2} G_F V_{us} f_+(0)~\bar{u}(p_{\nu})P_R \left[ A_{pole} \sla{p}_K
 - B_{pole} \sla{p}_{\ell}\ \right] v(p_{\ell}),
\end{equation}
where $A_{pole}$ and $B_{pole}$ are composed of loop integrals.  In
this section, we present the expressions for these parameters.  

We break-up the expression for $A_{pole}$ and for $B_{pole}$ into five
components,
\begin{eqnarray}
A_{pole} = \sum_{i=1}^{5} A_i & , & B_{pole} = \sum_{i=1}^{5} B_i,
\end{eqnarray}
where the `$1$' component is from inter-particle photon exchange, the
`$2$' component is from the $\pi$ and $\ell$ wavefunction
renormalizations, the `$3$' component is from photon exchange between
the effective vertex and the charged lepton, the `$4$' component is
from photon exchange between the effective vertex and the pion, and
the `$5$' component comes from emission and re-absorption of a photon
by the effective vertex.  Since single photon radiation from the
effective vertex is small, we expect that $A_5$ and $B_5$ (double
photon radiation) are negligible~\cite{footnote:doublerad}.

The expressions for the $A_i$ and the $B_i$ may be written, compactly,
in terms of a handful of two-, three-, and four- point functions.  We
define $B_i^{(\pi)} = B_i(\mpisq;\mpisq,\lambda^2)$ and $B_i^{(\ell)}
= B_i(\mlsq;\mlsq,\lambda^2)$ where $i=0,1$.  We also define the
two-point functions that depend on the pole masses $m_+$ and $m_0$,
\begin{equation}
\renewcommand\arraystretch{1.5}
\begin{array}{lcl}
B_0^{(\sigma,\,A)} & = & B_0(t;\lambda^2,m^2_{\sigma}) \\
B_0^{(\sigma,\,B)} & = & B_0(t;\mpisq,m^2_{\sigma}), \\
B_0^{(\sigma,\,C)} & = & B_0(\mksq;\mpisq,m^2_{\sigma}), 
\end{array}
\end{equation}
where $\sigma = +,\, 0$.  We define the following four-point functions:
\begin{equation}
\renewcommand\arraystretch{1.5}
\begin{array}{lcl}
C_i^{(\sigma,\,A)} &=& C_i(\mpisq,\mksq,t;\lambda^2,\mpisq,m^2_{\sigma})  \\
C_i^{(\sigma,\,B)} &=& C_i(s_{\pi\ell},0,\mksq;\mpisq,\mlsq,m^2_{\sigma}) \\
C_i^{(\sigma,\,C)} &=& C_i(\mlsq,t,0;\mlsq,\lambda^2,m^2_{\sigma}) \\
C_i^{(\sigma,\,D)} &=& C_i(\mpisq,t,\mksq;\mpisq,\lambda^2,m^2_{\sigma}),
\end{array}
\end{equation}
where $s_{ij}=(p_i-p_j)^2$, $i=0,1,2$ and $\sigma = +,\, 0$.  We define the
following four-point functions:

\begin{equation}
\renewcommand\arraystretch{1.5}
\begin{array}{l}
D_i^{(\sigma)} = D_i(\mpisq,s_{\pi\ell},0,t,\mlsq,\mksq;\lambda^2,\mpisq,\mlsq,m_{\sigma}^2),
\end{array}
\end{equation}
where $s_{ij}=(p_i-p_j)^2$, $i=0,1,2,3$ and $\sigma = +,\, 0$.

Using the above definitions, $A_1$ and $B_1$ are
\begin{eqnarray}\label{eqn:pole1AB}
A_1 & = & -2\mplussq C_0^{(+,\,A)}-2{\Delta C_0^B}-{\Delta C_1^A} 
          +8\mplussq x_{\pi\ell}\dzp -2\lambda^2{\Delta D_0} \nonumber \\
    & &   +4\mplussq\mpisq\dop +4x_{\pi\ell} \,{\Delta D_1} 
 +4\mplussq (\mlsq+2x_{\pi\ell} )\dtwp -2\mlsq \,{\Delta D_2} \nonumber \\
    & &   +4\mplussq (\mksq-\mpisq)\dthp - 2t \,{\Delta D_3}, \nonumber \\
B_1 & = & -\left[ \Delta C_0^{A} + \Delta C_1^{A} + \Delta C_2^{A} \right]
 -2\,{\Delta C_0^{B}} +[4x_{\pi\ell}-2\lambda^2]{\Delta D_0} +[4x_{\pi\ell}
 -2\mpisq]{\Delta D_1} \nonumber \\
    & + & 4\mplussq s_{\pi\ell}\,\dtwp + [4x_{\pi\ell}-2\mlsq]{\Delta D_2} 
          +4\mplussq\mksq\dthp +[4x_{\pi\ell}-2t]{\Delta D_3},
\end{eqnarray}
where 
\begin{equation}
\renewcommand\arraystretch{1.5}
\begin{array}{lcl}
\Delta C_i^{X} & = & \mplussq C_i^{(+,\,X)} 
                    +\mkmmpi \left[C_i^{(+,\,X)} - C_i^{(0,\,X)}\right], \\
\Delta D_i     & = & \mplussq \dip  
                    +\mkmmpi \left[\dip-\dio\right],
\end{array}
\end{equation}
and $X = A,\,B$.  The components $A_2$ and $B_2$ are expressed in
terms of the wavefunction renormalization constants [see
Eq.~(\ref{eqn:counterterms})],
\begin{equation}\label{eqn:pole2AB}
\renewcommand\arraystretch{1.5}
\begin{array}{lcl}
A_2 = \hat{f}_+(t)\left(\frac{4\pi}{\alpha}\right)
 \left( \delta Z_{\pi} + \delta Z_{\ell} \right) & , &
B_2 = \hat{f}_2(t)\left(\frac{2\pi}{\alpha}\right)
 \left( \delta Z_{\pi} + \delta Z_{\ell} \right),
\end{array}
\end{equation}
where $t=(p_K-p_{\pi})^2$.  The $A_i$ and $B_i$ components arising
from photon emission from the vertex and absorption by either the pion
or the lepton ($i=3,4$) are given by the equations
\begin{eqnarray}\label{eqn:pole3}
A_{3} & = & \; \hat{f}_+(t) \left[4\mlsq C_0^{(+,C)} - 4t C_1^{(+,C)}\right]
 \nonumber \\
B_{3} & = & \; \hat{f}_+(t)\left\{ -2B_0^{(0,\,A)} + \frac{2}{\mplussq}
 {\Delta B_0^{(A)}} - 8x_{K\nu} \left[ C_0^{(+,\,C)} + C_1^{(+,\,C)}
 + C_2^{(+,\,C)} \right] \right. \nonumber \\
 & -& \left. 2\,\mlsq \; \left[ C_0^{(0,\,C)} - m_+^{-2}{\Delta C_0^{(C)}}
 \right] + 2t \left[ C_1^{(0,\,C)} - m_+^{-2} {\Delta C_1^{(C)}} \right]
 \right\} \nonumber \\
 & & + \hat{f}_2(t)\left\{ 2B_0^{(0,\,A)} -4B_0^{(\ell)} - 2B_1^{(\ell)} +
 2\mlsq C_0^{(0,\,C)} - 2t C_1^{(0,\,C)} \right\},
\end{eqnarray}
and
\begin{eqnarray}\label{eqn:pole4}
A_{4} & = & \; \hat{f}_+(t) \left\{ 4B_0^{(A)} - 2B_0^{(C)} -2B_0^{(\pi)}
 -2\left[2(\mksq-\mpisq)-t+\lambda^2-\mplussq\right] C_0^{(+,\,D)} \right\}
 \nonumber \\
 & & +{\Delta C_0^{(D)}} - {\Delta C_1^{(D)}} - {\Delta C_2^{(D)}} \nonumber \\
B_{4} & = & \; {\Delta C_0^{(D)}} - {\Delta C_1^{(D)}} + \hat{f}_2(t)
 \left\{ 2B_0^{(+,\,A)} + B_0^{(\pi)} - B_0^{(+,\,C)} \right. \nonumber \\
 & & \left. - \left[ 2(\mksq-\mpisq)-t+\lambda^2-\mplussq \right] C_0^{(+,\,D)}
 \right\} \nonumber \\
 & & + \frac{2}{\mzerosq} \hat{f}_0(t)\left\{ -\mplussq B_0^{(+,\,A)}
  +\Delta B_0^{(A)} +\frac{1}{2}\mplussq B_0^{(+,\, B)}
  -\frac{1}{2} \Delta B_0^{(B)} \right. \nonumber \\
 & & -\frac{1}{2} \left[ 2(\mksq-\mpisq)-t+\lambda^2 \right] C_0^{(+,\, D)}
   \nonumber \\
 & & \left. + \frac{1}{2} \left[ 2(\mksq-\mpisq)-t+\lambda^2 \right]
 {\Delta C_0^{(D)}} + \mplussq C_0^{(+,\, D)} - \mzerosq C_0^{(0,\, D)}
 \right\}.
\end{eqnarray}

\begin{figure}[h]
\begin{center}
\includegraphics[scale=0.5]{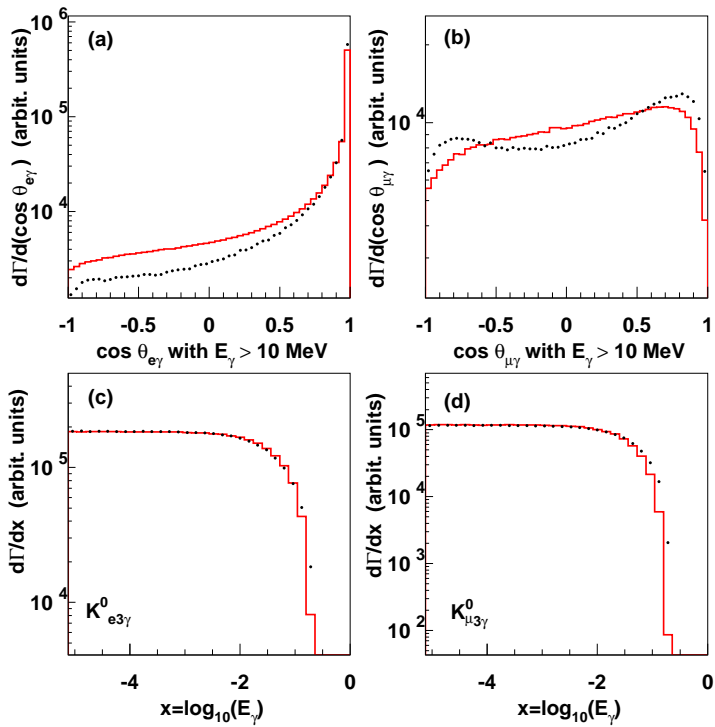}
\end{center}
\caption{\label{fig:photos} Comparison of the {\tt PHOTOS} (dots) and
the {\tt KLOR} (line) Monte Carlo generators.  (a) and (b) plot the
cosine of the angle between the charged lepton and the photon
($\cos\theta_{\ell\gamma}$) for the \ken and the \kmn decay modes,
respectively.  In plots (a) and (b) radiated photons are required to
have an energy greater than $10$ MeV in the kaon center of mass.
Plots (c) and (d) compare the log of the radiated photon energy
($\log_{10}(E_{\gamma})$), where $E_{\gamma}$ is in GeV.  In {\tt
PHOTOS} the IR cutoff is taken to be $1$ keV and in {\tt KLOR} the
photon mass was taken to be $1$ eV/c$^2$.}
\end{figure}

\section{Comparison of {\tt KLOR} to {\tt PHOTOS}}\label{appendix:c}
One tool often used by experimentalists to model radiative effects is
a ``universal'' Monte Carlo generator called {\tt
PHOTOS}~\cite{Barberio:1993qi}.  For the \kln decay mode, {\tt PHOTOS}
estimates the size of the radiative corrections in the
leading-logarithmic (collinear) approximation.  Since it treats
radiation from the charged lepton and the pion independently, {\tt
PHOTOS} should succeed in reproducing the photon energy distribution
but fail to reproduce distributions of the angle between the photon
and a charged particle.  To illustrate this behavior, we generate the
$\log(E_{\gamma})$ and the $\cos\theta_{\ell\gamma}$ distributions
using {\tt PHOTOS} and {\tt KLOR} [see Fig.~\ref{fig:photos}].  In
\ken decays, {\tt PHOTOS} significantly underestimates radiation in
the central and backward directions, while for \kmn decays, photon
radiation is suppressed in the central region.  For both the \ken and
\kmn decay modes, {\tt PHOTOS} and {\tt KLOR} agree for lower energy
photons.  The discrepancy at higher energies is most likely a result
of the ``soft photon'' assumption in {\tt PHOTOS}.

Though {\tt PHOTOS} is a powerful tool, precision measurements of \kln
decays require a more accurate understanding of radiative effects.
With {\tt KLOR}, we hope to satisfy this requirement and be able to
study the structure of the radiative corrections.  It should be noted
that the results of {\tt PHOTOS} can be improved if the \kln
inner-bremsstrahlung matrix element is incorporated into the {\tt
PHOTOS} algorithm.

% Bibliography %%%%%%%%%%%%%%%%%%%%%%%%%%%%%%%%%%%%%%%%%%%%%%%%%%%%%%


\begin{thebibliography}{99}
\bibitem{footnote:delta} The short- and long- distance corrections are
often written separately using the $S_{EW}$ and $\delta_K^{\ell}$
notation, respectively.  The two formulations are related by the
relation, $S_{EW}(1+\delta_K^{\ell}) = 1+\dtot = 1+\dlongl+\dshort$,
where $S_{EW}\simeq 1+\dshort$.
\bibitem{footnote:shortdistance} Numerical values for \dshort\spac are
obtained by using Eq.~(10) in Ref.~\cite{marciano}.  Eq.~(10) sums the
leading short-distance logs but does not include the QCD factor
$(1-\alpha_s/\pi)$.  Therefore, the calculated values of \dshort\spac
quoted in Table~\ref{tab:delta_ele} and Table~\ref{tab:delta_mu}
should be accurate to within $\sim 0.04\%$.
\bibitem{footnote:calculation_errors} The formulae for the radiative
correction in Ref.~\cite{Ginsberg:pz,Ginsberg:vy} are known to have
errors that affect the results.  The expressions have been updated in
two subsequent errata.  The numerical values of the long-distance
radiative corrections, $\hdlonge=1.5\%$ and $\hdlongm=2.0\%$, are from
the second errata of Ref.~\cite{Ginsberg:pz}.
\bibitem{footnote:doublerad} To calculate the effective vertex for
double-photon radiation one may proceed in an analogous manner to the
single-photon radiation case.  Modulo structure-dependent terms, this
vertex may be specified by requiring the Ward Identity to be
satisfied: $\epsilon^{\mu}(k_1)\epsilon^{\nu}(k_2){\mathcal
M}_{\mu\nu}$ is zero when either $\epsilon^{\mu}(k_1) \rightarrow
k^{\mu}_1$ or $\epsilon^{\nu}(k_2) \rightarrow k^{\nu}_2$.
\end{thebibliography}
\end{document}